\documentclass[cmp,final,10pt]{svjour}
\usepackage{amssymb}
\usepackage{amsfonts}
\usepackage{amsmath}
\usepackage{float}
\usepackage[all,web]{xy}

\newcommand{\Z}{\mathbb Z}
\newcommand{\R}{\mathbb R}
\newcommand{\C}{\mathbb R}
\newcommand{\+}{\!+\!}
\newcommand{\m}{\!-\!}
\newcommand{\half}{{\lower0.25ex\hbox{\raise.6ex\hbox{\the\scriptfont0 1}\kern-.2em\slash\kern-.1em\lower.25ex\hbox{\the\scriptfont0 2}}}}
\newcommand{\third}{{\lower0.25ex\hbox{\raise.6ex\hbox{\the\scriptfont0 1}\kern-.2em\slash\kern-.1em\lower.25ex\hbox{\the\scriptfont0 3}}}}
\newcommand{\thalf}{{\lower0.25ex\hbox{\raise.6ex\hbox{\the\scriptfont0 3}\kern-.2em\slash\kern-.1em\lower.25ex\hbox{\the\scriptfont0 2}}}}
\newcommand{\quarter}{{\lower0.25ex\hbox{\raise.6ex\hbox{\the\scriptfont0 1}\kern-.2em/\kern-.1em\lower.25ex\hbox{\the\scriptfont0 4}}}}
\newcommand{\eight}{{\lower0.25ex\hbox{\raise.6ex\hbox{\the\scriptfont0 1}\kern-.2em/\kern-.1em\lower.25ex\hbox{\the\scriptfont0 8}}}}

\begin{document}

\title{\large\bf
Application of the $\tau$-Function Theory of Painlev\'e Equations to Random 
Matrices: PIV, PII and the GUE}
\titlerunning{
Application of the $\tau$-Function Theory of PIV, PII to the GUE}

\date{Received: 27 June 2000 / Accepted: 8 December 2000}

\author{
P.J.~Forrester\inst{1} \and 
N.S.~Witte\inst{1,2}}
\institute{
Department of Mathematics and Statistics,
University of Melbourne, Victoria 3010, Australia.\\
\email{P.Forrester@ms.unimelb.edu.au}
\email{N.Witte@ms.unimelb.edu.au},
\and
School of Physics, University of Melbourne, Victoria 3010, Australia.}
\maketitle

\communicated{P. Sarnak}
\begin{abstract}

Tracy and Widom have evaluated the cumulative distribution of the largest
eigenvalue for the finite and scaled infinite GUE 
in terms of a PIV and PII transcendent respectively. We
generalise these results to the evaluation of
$\tilde{E}_N(\lambda;a) := 
\Big \langle \prod_{l=1}^N \chi_{(-\infty, \lambda]}^{(l)}
(\lambda - \lambda_l)^a \Big \rangle$,
where $ \chi_{(-\infty, \lambda]}^{(l)} = 1$ for
$\lambda_l \in (-\infty, \lambda]$ and $ \chi_{(-\infty, \lambda]}^{(l)} = 0$
otherwise, and the average is with respect to the joint eigenvalue
distribution of the GUE, as well as to the evaluation of
$F_N(\lambda;a) := \Big \langle \prod_{l=1}^N (\lambda - \lambda_l)^a
\Big \rangle$. Of particular interest are 
$\tilde{E}_N(\lambda;2)$ and $F_N(\lambda;2)$, and their scaled limits,
which give the distribution of the largest eigenvalue and the density
respectively. Our results are obtained by applying the Okamoto
$\tau$-function theory of PIV and PII, for which we give a self contained
presentation based on the recent work of Noumi and Yamada. We point
out that the same approach can be used to study the quantities
$\tilde{E}_N(\lambda;a)$ and $F_N(\lambda;a)$ for the other
classical matrix ensembles.

\end{abstract}

% \keywords{random matrices -- Painlev\'e equations -- Toda lattice equation
%  -- B\"acklund transformations -- root systems -- affine Weyl groups
%  -- classical orthogonal polynomial systems}
%
\tableofcontents

\section{Introduction and Summary}
\setcounter{equation}{0}
Hermitian random matrices $X$ with a unitary symmetry are defined so that
the joint distribution of the independent elements $P(X)$ is unchanged by
the similarity transformation  $X \mapsto U^\dagger X U$ for $U$ unitary.
For example, an ensemble of matrices with 
$ P(X) := \exp \sum^{\infty}_{j=0} \alpha_j {\rm Tr}(X^j)
        =: \prod^N_{j=1}g(\lambda_j) $ for
general $g(x) \ge 0$ possesses a unitary
symmetry. Such ensembles have the property that
the corresponding eigenvalue probability density function
$p(\lambda_1,\dots,\lambda_N)$ is given by the explicit functional
form
\begin{equation}\label{1f}
p(\lambda_1,\dots,\lambda_N) = {1 \over C} \prod_{l=1}^N
g(\lambda_l) \prod_{1 \le j < k \le N} |\lambda_k - \lambda_j|^2,
\end{equation}
$C$ denoting the normalization.
(Throughout the symbol $C$ will be used to denote {\it some}
constant, i.e.~a quantity independent of the primary variable(s)
of the equation.) The choice $g(x) = e^{-x^2}$, which is
realized by choosing each diagonal element of $X$ independently from
the normal distribution N$[0,1/\sqrt{2}]$, and each off diagonal
independently with distribution N$[0,1/2] + i {\rm N}[0,1/{2}]$,
is referred to as the Gaussian Unitary Ensemble (GUE) and is the main
focus of the present article. Specifically our interest is in the
distribution of the largest eigenvalue, and the average values of powers 
(integer and fractional) of the
characteristic polynomial
$
\prod_{l=1}^N (\lambda - \lambda_l)
$
for such matrices.

Let $E_N(0;(s,\infty))$ denote the probability that there are no
eigenvalues in the interval $(s,\infty)$ for $N \times N$ GUE matrices.
The distribution of the largest eigenvalue $p_{\rm max}(s)$ is given
in terms of $E_N(0;(s,\infty))$ by
\begin{equation}\label{1.1'}
p_{\rm max}(s) = {d \over ds} E_N(0;(s,\infty)).
\end{equation}
With $R_N(s)$ specified by the solution of the nonlinear equation
\begin{equation}\label{c1a}
(R_N'')^2 + 4(R_N')^2(R_N' + 2N) - 4(sR_N' - R_N)^2 = 0,
\end{equation}
(an example of the Jimbo-Miwa-Okamoto $\sigma$-form of the
Painlev\'e IV differential equation; see Eq. (\ref{f.5}) below)
subject to the boundary condition
\begin{equation}
R_N(s) \mathop{\sim}\limits_{s \to \infty} {2^{N-1} s^{2N-2} e^{-s^2}
\over \pi^{1/2} (N\m 1)!},
\end{equation}
it has been shown by Tracy and Widom \cite{TW-94} that
\begin{equation}\label{2}
E_N(0;(s,\infty)) = \exp\Big( -\int_s^\infty R_N(t) \, dt \Big).
\end{equation}
The derivation in \cite{TW-94} uses functional properties of Fredholm
determinants  (a subsequent derivation using the KP equations and
Virasoro algebras has been given by Adler et al.~\cite{ASV-95}). In this
work we will give a derivation of (\ref{2}) based on the $\tau$-function
theory of the Painlev\'e IV equation due to Okamoto \cite{Ok-86}, and refined
by Noumi and Yamada \cite{NY-97}. Our principal observation in employing
this body of theory to problems in random matrix theory is that there is a
deep and fundamental relationship between 
$\tau$-functions relating to the Hamiltonian formalism of the
Painlev{\'e} theory, and particular multiple integrals specifying averages
with respect to the probability density function (\ref{1f}) in the
case that $g(x)$ takes the form of a classical weight function. From
the random matrix perspective the classical weights are \cite{F-00}
\begin{equation}
  g(x) = \left \{ \begin{array}{ll}
  e^{-x^2}, & {\rm Gaussian} \\
  x^a e^{-x} \: (x>0), & {\rm Laguerre} \\
  (1-x)^a(1+x)^b \: (-1<x<1), & {\rm Jacobi} \\
  (1+x^2)^{-\alpha}, & {\rm Cauchy} \ . \end{array} \right.
\label{weights}
\end{equation} 
To summarise the correspondence, which applies to any of the
cases, we set out the following schematic table:

\begin{table}[H]
\begin{center}
\begin{tabular}{|c|c|}
\hline

Painlev{\'e} Theory & Random Matrix Ensembles \\
\hline

$ \tau$-function $ \tau[N](s;a) $ &
\begin{minipage}[c]{7cm}
  Gap probability $ E_N(s) $ \\
  Averages $ \tilde{E}_N(s;a), F_N(s;a) $ \\
\end{minipage} \\
\hline

Hamiltonian $ H[N](s;a) $ &
\begin{minipage}[c]{7cm}
Resolvent kernel $ R_N(s;a)$ \\
Logarithmic derivative of averages - \\
$ U_N(s;a), V_N(s;a) $ 
\end{minipage} \\
\hline

Classical solutions - Weyl chamber walls &
Classical weights - Determinant structure \\
\hline

\end{tabular}
\caption{Correspondence between random matrix theory and Painlev{\'e} theory.}
\label{PJ-RMT}
\end{center}
\end{table}

The quantity $ \tilde{E}_N(s;a)$ in Table 1 is specified by
\begin{equation}\label{1.3'}
\tilde{E}_N(\lambda;a) := 
	\Big \langle \prod_{l=1}^N \chi_{(-\infty, \lambda]}^{(l)}
	(\lambda - \lambda_l)^a \Big \rangle, 
\end{equation}
where $ \chi_{(-\infty, \lambda]}^{(l)} = 1$ for
$\lambda_l \in (-\infty, \lambda]$ and $ \chi_{(-\infty, \lambda]}^{(l)} = 0$
otherwise, and the
 average is with respect to the eigenvalue probability density
function (\ref{1f}). For general $a$ we obtain the evaluation
\begin{equation}\label{1.6a}
 \tilde{E}_N(s;a) =  
 \tilde{E}_N(s_0;a) \exp \Big( \int_{s_0}^s U_N(t;a) \; dt \Big)
\end{equation}
(Eq.~(\ref{ent4}) with the substitution (\ref{UH})), where $U_N(t;a)$
satisfies the nonlinear equation
\begin{equation}
(U_N'')^2 - 4(tU_N' - U_N)^2 + 4U_N'(U_N' - 2a)(U_N'+2N) = 0,
\end{equation}
(Eq.~(\ref{rsw})) subject to the boundary condition
\begin{equation}
U_N(t;a) \mathop{\sim}\limits_{t \to
-\infty} - 2N t - {N(a+N) \over t}  + O\Big ( {1 \over t^3} \Big ),
\end{equation}
(Eq.~(\ref{rsw6})). For $(N\+ 1) \times (N\+ 1)$ dimensional GUE matrices
$p_{\rm max}(s)$ is proportional to $e^{-s^2} \tilde{E}_N(s;2)$.
We therefore have
\begin{equation}
p_{\rm max}(s) \Big |_{N \mapsto N +1} =
p_{\rm max}(s_0) \Big |_{N \mapsto N +1}
 \exp \Big( \int_{s_0}^s\left[ -2t + U_N(t;2) \right]\; dt \Big),
\end{equation}
(Eq.~(\ref{ag})).

The quantity $F_N(s;a)$ in Table 1 is specified by
\begin{equation}\label{FN}
F_N(\lambda;a) := \Big \langle \prod_{l=1}^N (\lambda - \lambda_l)^a
		  \Big \rangle \ ,
\end{equation}
(Eq.~(\ref{1.3'})).
For general positive integers $a$
(\ref{FN}) has been computed by Br\'ezin and Hikami in terms of the
determinant of an $a \times a$ matrix involving Hermite polynomials.
Note that for $a$ not equal to a positive integer, (\ref{FN}) is well
defined provided $\lambda$ has a non-zero imaginary part.
For general $a$ we obtain the evaluation
\begin{equation}
F_N(\lambda;a) = F_N(\lambda_0;a)
        \exp \Big( \int_{\lambda_0}^\lambda V_N(t;a) \; dt \Big) \ ,
\end{equation}
(Eq.~(\ref{bf1})) where $V_N(t;a)$ also satisfies the nonlinear equation 
(\ref{rsw}), but now with the boundary conditions
\begin{equation}
V_N(t;a) \mathop{\sim}\limits_{t \to \pm \infty} \chi {N a \over t}
\Big (1 + O(1/t) \Big ) \quad {\rm as} \quad t \to \infty
\end{equation}
(Eq.~(\ref{rs3})) where $\chi = 1$ for $t \to \infty$ and $| \chi | =1$ 
for $t \to - \infty$.
In the case $a=2$ this average is proportional to the
polynomial part of the eigenvalue density for $(N\+ 1) \times (N\+ 1)$
dimensional GUE matrices, which in terms of the Hermite polynomial
$H_N(\lambda)$ is proportional to each of the $2 \times 2$ determinants
termed Tur\'anians \cite{KS-82}
\begin{equation}\label{1.1}
\left [ \begin{array}{cc} H_{N}(\lambda) & 
H_{N+1}(\lambda) \\
H_{N}'(\lambda) & H_{N+1}'(\lambda) \end{array} \right ], \quad
\left [ \begin{array}{cc} H_{N+1}(\lambda) &
H_{N+1}'(\lambda) \\
H_{N+1}'(\lambda) & H_{N+1}''(\lambda) \end{array} \right ], \quad
\left [ \begin{array}{cc} H_{N}(\lambda) &
H_{N+1}(\lambda) \\
H_{N+1}(\lambda) & H_{N+2}(\lambda) \end{array} \right ] \ ,
\end{equation}
(which are of course proportional to each other).
The result (\ref{bf1}) with $a=2$ implies
\begin{equation}
\rho(\lambda) \Big |_{N \mapsto N + 1} =
\rho(\lambda_0) \Big |_{N \mapsto N + 1}
\exp \Big( \int_{\lambda_0}^\lambda \left[ -2t + V_N(t;2) \right]\; dt \Big),
\end{equation}
(Eq.~(\ref{bf7})).

In Sect. 2 we review the $\tau$-function theory of the Painlev\'e IV
equation, revising relevant aspects of the work of Okamoto 
\cite{Ok-81,Ok-86},
Noumi and Yamada \cite{NY-97,NY-2000} and Kajiwara et al.~\cite{KMNOY-99}. 
The culmination of this theory from our perspective is 
the derivation of determinant formula
expressions for the $\tau$-function corresponding to special values of
the parameters in the Painlev\'e IV equation. On the other hand,
it follows easily from the definitions that $\tilde{E}_N$ and $F_N$ can be
written as determinants. These are presented in Sect. 4. The
determinant formulas in fact precisely coincide with those occurring in
Sect. 2, so consequently we can characterise both $ \tilde{E}_N $ and $ F_N $
in terms of solutions of the nonlinear equation (\ref{rsw}). 
The theory presented in Sect. 2 also allows $ \tilde{E}_N $, $ F_N $
to be characterised as solutions of a certain fourth order difference
equation (Eq.~(\ref{taudiff-T3})), and $ U_N $, $ V_N $ as solutions of a
particular third order difference equation (\ref{Udiff-T3}).

Also of interest is the scaling limit of (\ref{1.3'}) and (\ref{FN})
with $\lambda \mapsto \sqrt{2N} + \lambda/\sqrt{2} N^{1/6}$. This choice
of coordinate corresponds to shifting the origin to the edge of the leading
order support of the eigenvalue density, then scaling the coordinate so
as to make the spacings of order unity as $N \to \infty$. We find the
scaled quantities can be expressed in terms of particular solutions
of the general Jimbo-Miwa-Okamoto $\sigma$ form of the Painlev\'e
II equation
\begin{equation}
(u'')^2 + 4 u' \Big ( (u')^2 - s u' + u \Big ) - a^2 = 0 \ ,
\end{equation}
(Eq.~(\ref{4.4})). Specifically, as already known from \cite{TW-94a},
\begin{equation}
E^{\rm soft}(s) := \lim_{N \to \infty} E_N\Big (0;
\sqrt{2N} + {s \over \sqrt{2} N^{1/6}} \Big ) 
= \exp \Big( -\int_s^\infty r(t) \; dt \Big) \ ,
\end{equation}
where $r(s)$ satisfies (\ref{4.4}) with $a=0$. Also
\begin{equation}
\tilde{E}^{\rm soft}(s;a) :=
\lim_{N \to \infty \atop
s \mapsto \sqrt{2N} + s/\sqrt{2} N^{1/6}} 
\Big ( C e^{-as^2/2} \tilde{E}_N(s;a) \Big )
=
\tilde{E}^{\rm soft}(s_0;a)
\exp \Big( \int_{s_0}^s u(t;a) \; dt \Big) \ ,
\end{equation}
where $u(s;a)$ satisfies (\ref{4.4}) subject to the
boundary condition 
\begin{equation}
   u(s;a) \mathop{\sim}\limits_{s \to - \infty}
   \quarter s^2 + {4a^2\m 1 \over 8s} + {(4a^2\m 1)(4a^2\m 9) \over 64s^4}
   + \ldots \ ,
\end{equation}
(Eq.~(\ref{es-bcU})). In the case $a=2$ (Eq.~(\ref{4.3b})) gives the formula
\begin{equation}
p_{\rm max}^{\rm soft}(s) =
p_{\rm max}^{\rm soft}(s_0) \exp \Big( \int_{s_0}^s u(t;2) \; dt \Big),
\end{equation}
(Eq.~({\ref{pms})) for the scaled distribution of the largest eigenvalue 
in the GUE.

Analogous to the formula (\ref{4.3b}), for the scaled limit of
$F_N(\lambda;a)$ we have
\begin{equation}
{F}^{\rm soft}(\lambda;a) :=
\lim_{N \to \infty \atop
\lambda \mapsto \sqrt{2N} + \lambda/\sqrt{2} N^{1/6}}
\Big ( C e^{-a\lambda^2/2} {F}_N(\lambda;a) \Big ) =
{F}^{\rm soft}(\lambda_0;a)
\exp \Big( \int_{\lambda_0}^\lambda v(t;a) \; dt \Big) \ ,
\end{equation}
(Eq.~({\ref{4.4b})) where $v(s;a)$, like $u(s;a)$, satisfies (\ref{4.4}). 
The difference between
$u$ and $v$ is in the boundary condition; for the latter we require
\begin{equation}
  v(t;a)  \mathop{\sim}\limits_{t \to \infty}
  -a t^{1/2} - {a^2 \over 4t} + {a(4a^2\+ 1) \over 32t^{5/2}} \ 
\end{equation}
(Eq.~(\ref{air-v})). The case $a=2$ corresponds to the scaled eigenvalue
density at the spectrum edge, which has the known evaluation
\cite{Fo-93}
\begin{equation}
\rho^{\rm soft}(s) =
- \left| \begin{array}{cc} {\rm Ai}(s) & {\rm Ai}'(s) \\
                           {\rm  Ai}'(s) & {\rm Ai}''(s) \end{array} \right| ,
\end{equation}
where Ai$(s)$ denotes the Airy function. In fact for all $a \in \Z_{\ge 0}$
we have the determinantal form
\begin{equation}
F^{\rm soft}(\lambda;a) = (-1)^{a(a-1)/2}
\det \Big [ {d^{j+k} \over d\lambda^{j+k}} {\rm Ai} \, (\lambda)
\Big ]_{j,k=0,\dots,a-1} \ ,
\end{equation}
(Eq.~({\ref{air})).

In Sect. 3 we present the $\tau$-function theory of the Painlev\'e II
equation in an analogous fashion to the theory presented in Sect.
2 for the  Painlev\'e IV equation. In particular we derive the
second order second degree equation satisfied by the Hamiltonian
(which is known from \cite{JM-81} and \cite{Ok-86}) as well as a fourth
order difference equation satisfied by the $\tau$-functions. Also
derived is the fact that the right-hand side of (\ref{air})
corresponds to a $\tau$-function sequence in the PII theory, which is
a result of Okamoto \cite{Ok-86}. In Sect. 5 the results (\ref{4.1a}),
(\ref{4.3b}), (\ref{4.4b}) and (\ref{air}) are derived from a limiting
process applied to the corresponding finite $N$ results.

A programme for further study is outlined in Sect. 6.

\section{$\tau$-Function Theory for PIV}
\setcounter{equation}{0}
\subsection{Affine Weyl group symmetry}
It has been demonstrated in the works of Okamoto \cite{Ok-86} (in a series
of papers treating all the Painlev{\'e} equations), Noumi
and Yamada \cite{NY-97} (see also their works \cite{NY-98,NY-98a,NY-2000})
and the earlier work of Adler \cite{Ad-94}
that the fourth Painlev{\'e} equation
\begin{equation}\label{PIV}
 y'' = {1 \over 2y} (y')^2 + {3 \over 2} y^3 + 4t y^2 +
	2(t^2 - \alpha) y + {\beta \over y} \ ,
\end{equation}
can be recast in a way which reveals its symmetries in a particularly
manifest and transparent form. 

\begin{proposition}[\cite{NY-97,NY-2000}]
The fourth Painlev{\'e} equation is equivalent to the coupled set of autonomous
differential equations (where $' = d/dt$) 
\begin{equation}
\begin{split}
   f_0' & = f_0(f_1 - f_2) + 2 \alpha_0 \ ,\\
   f_1' & = f_1(f_2 - f_0) + 2 \alpha_1 \ ,\\
   f_2' & = f_2(f_0 - f_1) + 2 \alpha_2 \ ,
\end{split}
\label{c1}
\end{equation}
with $y= - f_1$ and where the parameters $ \alpha_j \in \C $ with 
$ \alpha_0 + \alpha_1 + \alpha_2 = 1 $ are related by 
\begin{equation}\label{PIVa}
\alpha = \alpha_0 - \alpha_2, \qquad \beta = - 2\alpha_1^2 \ ,
\end{equation}
and the constraint taken conventionally as
\begin{equation}\label{c2}
f_0 + f_1 + f_2 = 2t \ . 
\end{equation}
\end{proposition}
\noindent {\it Proof}.
Equation (\ref{c2}) reduces the three first order equations of
(\ref{c1}) down to two. Eliminating a further variable by introducing a
second derivative shows that $y= - f_1$ satisfies the PIV equation.
The form of these equations implies
\begin{equation}
(f_0 + f_1 + f_2)' = 2\alpha_0 + 2 \alpha_1 + 2 \alpha_2 = k,
\end{equation}
$ k \neq 0 $ constant, thus permitting the normalization given above.
\qed

\noindent {\it Note}.
Many differing conventions are in use for such a description of the PIV 
system and for example we have 
written $2 \alpha_j$ ($j=0,1,2$) in place of the $\alpha_j$ used in 
\cite{NY-97,NY-98a,NY-2000} in order to eliminate unnecessary factors of two 
appearing in the ensuing theory.

The hyperplane $\alpha_0 + \alpha_1 + \alpha_2 = 1$ in parameter space
$ (\alpha_0, \alpha_1, \alpha_2) \in \C^3 $ is associated with the simple
roots $ \alpha_0, \alpha_1, \alpha_2 $ spanning the root
system of type $A^{(1)}_2$. From this perspective the parameters 
$\alpha_0, \alpha_1$ and $\alpha_2$ define a triangular lattice in the 
plane (see Fig. \ref{A2-fig}). 

% \documentclass{article}
% \usepackage[all,web]{xy}
% \begin{document}
% \pagestyle{empty}

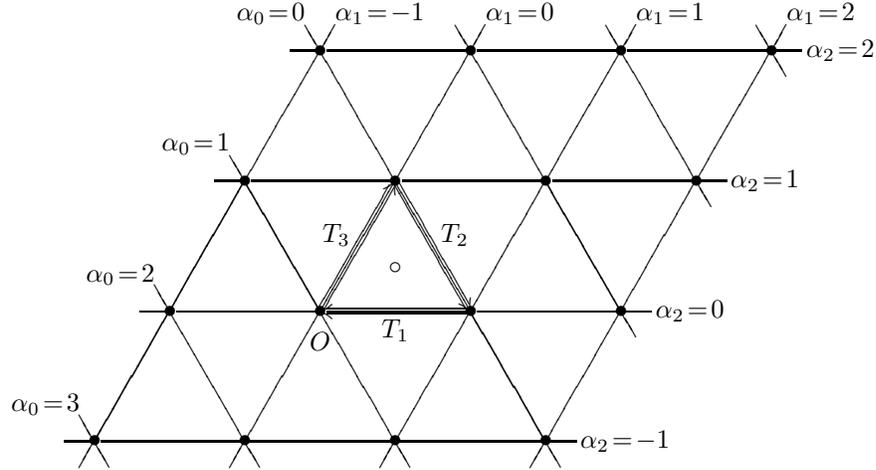
\begin{figure}[H]
\begin{center}

  \newcommand{\triad}{
   \save 
  {0*{\bullet};<2cm,0cm>**@{-}
   ,0;<+1.0cm,1.73205081cm>**@{-}
   ,0;<-1.0cm,1.73205081cm>**@{-}
  }\restore}

\[
 \renewcommand{\latticebody}{\drop{\triad}}
 \begin{xy}
    *\xybox{0;<2.0cm,0cm>:<1.0cm,1.73205081cm>::
           ,0,{\xylattice{0}{1}{0}{1}},
	   ,0+< 0cm,-.4cm>*{O}
	   ,0+< 1cm,0.57735027cm>*{\circ}
	   ,{\ar@{=>}^{\displaystyle T_1} 0+(1,0);0}
	   ,{\ar@{=>}^{\displaystyle T_3} 0;0+(0,1)}
	   ,{\ar@{=>}^{\displaystyle T_2} 0+(0,1);0+(1,0)}
%	   ,{\ar@*{[|(2)]}^{\displaystyle T_1} 0+(1,0);0}
%	   ,{\ar@*{[|(2)]}^{\displaystyle T_3} 0;0+(0,1)}
%	   ,{\ar@*{[|(2)]}^{\displaystyle T_2} 0+(0,1);0+(1,0)}
	   ,0+(-1,-1)="O"*{\bullet};"O"+<2.0cm,0cm>**@{-}
 	   ,"O";"O"+< 1cm,1.73205081cm>**@{-}
 	   ,"O";"O"+<-.2cm,-.34641016cm>**@{-}
 	   ,"O";"O"+< .2cm,-.34641016cm>**@{-}
 	   ,"O";"O"+<-.4cm,0cm>**@{-}
	   ,"O";"O"+<-.2cm, .34641016cm>**@{-}*!RD{\alpha_0\!=\!3}
	   ,0+(-1,0)="O"*{\bullet};"O"+<2.0cm,0cm>**@{-}
 	   ,"O";"O"+< 1cm,1.73205081cm>**@{-}
 	   ,"O";"O"+<-1cm,-1.73205081cm>**@{-}
 	   ,"O";"O"+< 1cm,-1.73205081cm>**@{-}
 	   ,"O";"O"+<-.4cm,0cm>**@{-}
	   ,"O";"O"+<-.2cm, .34641016cm>**@{-}*!RD{\alpha_0\!=\!2}
	   ,0+(-1,1)="O"*{\bullet};"O"+<2.0cm,0cm>**@{-}
 	   ,"O";"O"+< 1cm,1.73205081cm>**@{-}
 	   ,"O";"O"+<-1cm,-1.73205081cm>**@{-}
 	   ,"O";"O"+< 1cm,-1.73205081cm>**@{-}
 	   ,"O";"O"+<-.4cm,0cm>**@{-}
	   ,"O";"O"+<-.2cm, .34641016cm>**@{-}*!RD{\alpha_0\!=\!1}
           ,0+(-1,2)="O"*{\bullet};"O"+<2.0cm,0cm>**@{-}
 	   ,"O";"O"+< .2cm, .34641016cm>**@{-}*!LD{\alpha_1\!=\!-1}
 	   ,"O";"O"+<-.4cm,0cm>**@{-}
	   ,"O";"O"+<-.2cm, .34641016cm>**@{-}*!RD{\alpha_0\!=\!0}
           ,0+( 0,2)="O"*{\bullet};"O"+<2.0cm,0cm>**@{-}
 	   ,"O";"O"+<-.2cm, .34641016cm>**@{-}
 	   ,"O";"O"+< .2cm, .34641016cm>**@{-}*!LD{\alpha_1\!=\!0}
           ,0+( 1,2)="O"*{\bullet};"O"+<2.0cm,0cm>**@{-}
 	   ,"O";"O"+<-.2cm, .34641016cm>**@{-}
 	   ,"O";"O"+< .2cm, .34641016cm>**@{-}*!LD{\alpha_1\!=\!1}
           ,0+( 2,2)="O"*{\bullet}
 	   ,"O";"O"+<-.2cm, .34641016cm>**@{-}
 	   ,"O";"O"+< .2cm, .34641016cm>**@{-}*!LD{\alpha_1\!=\!2}
 	   ,"O";"O"+< .2cm,-.34641016cm>**@{-}
 	   ,"O";"O"+< .4cm, .0cm>**@{-}*!L{\,\alpha_2\!=\!2}
           ,"O";"O"+<-1cm,-1.73205081cm>**@{-}
           ,0+( 2,1)="O"*{\bullet}
 	   ,"O";"O"+< .2cm,-.34641016cm>**@{-}
 	   ,"O";"O"+< .4cm, .0cm>**@{-}*!L{\,\alpha_2\!=\!1}
           ,"O";"O"+<-1cm,-1.73205081cm>**@{-}
           ,"O";"O"+<-1cm, 1.73205081cm>**@{-}
           ,0+( 2,0)="O"*{\bullet}
 	   ,"O";"O"+< .2cm,-.34641016cm>**@{-}
 	   ,"O";"O"+< .4cm, .0cm>**@{-}*!L{\,\alpha_2\!=\!0}
           ,"O";"O"+<-1cm,-1.73205081cm>**@{-}
           ,"O";"O"+<-1cm, 1.73205081cm>**@{-}
           ,"O";"O"+<-1cm, 1.73205081cm>**@{-}
           ,0+( 2,-1)="O"*{\bullet}
 	   ,"O";"O"+< .2cm,-.34641016cm>**@{-}
 	   ,"O";"O"+<-.2cm,-.34641016cm>**@{-}
 	   ,"O";"O"+< .4cm, .0cm>**@{-}*!L{\,\alpha_2\!=\!-1}
           ,"O";"O"+<-2cm,0.0cm>**@{-}
           ,"O";"O"+<-1cm, 1.73205081cm>**@{-}
           ,"O";"O"+<-1cm, 1.73205081cm>**@{-}
           ,0+( 1,-1)="O"*{\bullet}
 	   ,"O";"O"+< .2cm,-.34641016cm>**@{-}
 	   ,"O";"O"+<-.2cm,-.34641016cm>**@{-}
           ,"O";"O"+<-2cm,0.0cm>**@{-}
           ,"O";"O"+< 1cm, 1.73205081cm>**@{-}
           ,"O";"O"+<-1cm, 1.73205081cm>**@{-}
           ,0+( 0,-1)="O"*{\bullet}
 	   ,"O";"O"+< .2cm,-.34641016cm>**@{-}
 	   ,"O";"O"+<-.2cm,-.34641016cm>**@{-}
           ,"O";"O"+< 1cm, 1.73205081cm>**@{-}
           }
 \end{xy} \]

\caption{Parameter space for $ (\alpha_0, \alpha_1, \alpha_2) $ associated
with the simple roots of the root system $ A_2^{(1)} $.}\label{A2-fig}
\end{center}
\end{figure}

% \end{document}

Let the fundamental reflections $s_i$ $(i=0,1,2)$ represent the automorphism of
the lattice specified by a reflection with respect to the line $\alpha_i = 0$.
Their action on the simple roots are given by
\begin{equation}\label{s1}
s_i(\alpha_j) = \alpha_j - \alpha_i a_{ij} \ ,
\end{equation}
where $a_{ij}$ are the elements of the Cartan matrix
\begin{equation}
A = \left [ \begin{array}{rrr}2&-1&-1\\-1&2&-1\\-1&-1&2\end{array} \right ].
\end{equation}
Let $\pi$ represent the lattice automorphism corresponding to a 
rotation by $120^\circ$ degrees around the barycentre of the 
fundamental alcove $C$ defined by $\alpha_i > 0$ $(i=0,1,2)$.
Then
\begin{equation}\label{s2}
\pi(\alpha_j) = \alpha_{j+1} \ ,
\end{equation}
$ j \in \Z/3\Z $. 
The operators $\pi$, $s_i$ obey the algebra
\begin{equation}
s_j^2 = 1, \quad (s_j s_{j+1})^3 = 1, 
\quad s_js_{j\pm 1}s_j = s_{j\pm 1}s_js_{j\pm 1},
\quad \pi^3 = 1, \quad \pi s_j = s_{j+1} \pi \ ,
\end{equation}
and generate $\widetilde{W} = \langle \pi, s_0, s_1, s_2 \rangle$ defining an 
extension of the affine Weyl group associated with the $A^{(1)}_2$ root system.

\begin{proposition}[\cite{NY-98,NY-98a}]
The B{\"a}cklund transformations of the PIV system are given by the actions of 
the extended affine Weyl group $\widetilde{W}$ on the parameters as specified
by (\ref{s1}) and (\ref{s2}), and on the functions as specified by
\begin{equation}\label{s3}
s_i(f_j) = f_j + {2 \alpha_i \over f_i} u_{ij} \ ,
\qquad \pi(f_j) = f_{j+1} \quad (i,j=0,1,2) \ , 
\end{equation}
where the $u_{ij}$ are the elements of the orientation matrix
\begin{equation}
U = \left[ \begin{array}{rrr}0&1&-1\\-1&0&1\\1&-1&0\end{array} \right] \ ,
\end{equation}
associated with the boundary of the fundamental alcove \cite{NY-97}.
\end{proposition}
\noindent {\it Proof}.
Let $V$ denote one of $\pi, s_0, s_1, s_2$ and let $\beta_i := V(\alpha_i)$.
Using (\ref{s1}) and (\ref{s2}) it's a simple exercise to explicitly
verify that $ g_i:= V(f_i) $ of the form (\ref{s3}) satisfy the structurally 
identical equations
\begin{equation}
\begin{split}
   g_0' & = g_0(g_1 - g_2) + 2 \beta_0 \ , \\
   g_1' & = g_1(g_2 - g_0) + 2 \beta_1 \ , \\
   g_2' & = g_2(g_0 - g_1) + 2 \beta_2 \ ,
\end{split}
\label{c1'}
\end{equation}
thus giving rise to the stated B\"acklund transformation. 
\qed

Following \cite{NY-2000,KMNOY-99}, in Tables \ref{A2-param}, \ref{A2-Bxfm} 
the actions (\ref{s1}), (\ref{s2}) and (\ref{s3}), are listed in tabular format.

\begin{table}[H]
\begin{center}
\begin{tabular}{|c|c|c|c|}
\hline                           

& $ \alpha_0 $ & $ \alpha_1 $ & $ \alpha_2 $ \\     
\hline                                                                     

  $ s_0 $
& $ -\alpha_0 $         
& $ \alpha_0 + \alpha_1 $
& $ \alpha_0 + \alpha_2 $ \\
\hline                 

  $ s_1 $               
& $ \alpha_1 + \alpha_0 $
& $ -\alpha_1 $
& $ \alpha_1 + \alpha_2 $ \\
\hline                

  $ s_2 $               
& $ \alpha_2 + \alpha_0 $
& $ \alpha_2 + \alpha_1 $
& $ -\alpha_2 $ \\
\hline                

  $ \pi $
& $ \alpha_1 $                                                                  
& $ \alpha_2 $                                                                  
& $ \alpha_0 $ \\                                         
\hline                                                                          
\hline                                                                          

  $ T_1 $
& $ \alpha_0 + 1 $
& $ \alpha_1 - 1 $
& $ \alpha_2 $ \\
\hline

  $ T_2 $
& $ \alpha_0 $
& $ \alpha_1 + 1 $
& $ \alpha_2 - 1 $ \\
\hline

  $ T_3 $
& $ \alpha_0 - 1 $
& $ \alpha_1 $
& $ \alpha_2 + 1 $ \\
\hline

\end{tabular}                                                                   
\caption{Action of the generators of the extended affine Weyl group associated
with the root system $ A^{(1)}_2 $ on the simple roots.}\label{A2-param}
\end{center}
\end{table}

\begin{table}[H]
\begin{center}
\begin{tabular}{|c|c|c|c|}
\hline

& $ f_0 $ & $ f_1 $ & $ f_2 $ \\
\hline

  $ s_0 $
& $ f_0 $
& \begin{minipage}[c]{2cm}{
  \begin{equation*}
     f_1 + {2\alpha_0 \over f_0}
  \end{equation*}
  }\end{minipage}
& \begin{minipage}[c]{2cm}{
  \begin{equation*}
     f_2 - {2\alpha_0 \over f_0}
  \end{equation*}
  }\end{minipage} \\
\hline

  $ s_1 $
& \begin{minipage}[c]{2cm}{
  \begin{equation*}
     f_0 - {2\alpha_1 \over f_1}
  \end{equation*}
  }\end{minipage}
& $ f_1 $
& \begin{minipage}[c]{2cm}{
  \begin{equation*}
     f_2 + {2\alpha_1 \over f_1}
  \end{equation*}
  }\end{minipage} \\
\hline

  $ s_2 $
& \begin{minipage}[c]{2cm}{
  \begin{equation*}
     f_0 + {2\alpha_2 \over f_2}
  \end{equation*}
  }\end{minipage}
& \begin{minipage}[c]{2cm}{
  \begin{equation*}
     f_1 - {2\alpha_2 \over f_2}
  \end{equation*}
  }\end{minipage}
& $ f_2 $ \\
\hline

  $ \pi $
& $ f_1 $
& $ f_2 $
& $ f_0 $ \\
\hline

\end{tabular}
\caption{B{\"a}cklund transformations for the PIV system.}\label{A2-Bxfm}
\end{center}
\end{table}

From the earlier work of Okamoto it has been known that the PIV system,
as for all the Painlev\'e transcendents, admits a Hamiltonian formulation
and that from this viewpoint the B{\"a}cklund transformations are 
birational canonical transformations 
$ \{q,p;H\} \mapsto \{\tilde{q},\tilde{p};\tilde{H}\} $.

\begin{proposition}[\cite{Ok-86,KMNOY-99}]
The PIV dynamical system is a Hamiltonian system $ \{q,p;H\} $ with the
Hamiltonian
\begin{equation}
\begin{split}
H  & = (2p - q - 2t)pq - 2 \alpha_1 p - \alpha_2 q \ ,\\
   & = \half f_0 f_1 f_2 + \alpha_2 f_1 - \alpha_1 f_2 \ ,
\end{split}
\label{7.a}
\end{equation}
and canonical variables $ q, p $
\begin{equation}\label{12'}
- f_1 = q, \qquad f_2 = 2p \ .
\end{equation}
\end{proposition}
\noindent {\it Proof}.
With $H$ specified by (\ref{7.a}), Hamilton's equations of motion read
\begin{equation}\label{7.1}
q' = {\partial H \over \partial p} = q(4p-q-2t)-2\alpha_1 \ ,
\qquad
p' = - {\partial H \over \partial q} = p(2q-2p+2t)+\alpha_2 \ .
\end{equation}
Substituting for $p$ and $q$ according to (\ref{12'}) shows
that these equations are identical to the final two equations in
(\ref{c1}). 
\qed

\noindent {\it Note}.
Because $-f_1$ satisfies the PIV equation (\ref{PIV}), it follows
immediately from the first equation in
(\ref{12'}) that $q$ satisfies the PIV equation (\ref{PIV}). 
Furthermore, use of the first equation in
(\ref{7.1}) shows 
\begin{equation}\label{2.11'}
p = {1 \over 4q}(q' + q^2 + 2tq + 2 \alpha_1) \ ,
\end{equation}
so $H$ is completely specified in terms of the Painlev\'e IV
transcendent (\ref{PIV}) with parameters (\ref{PIVa}).

There is a degree of ambiguity in constructing a Hamiltonian in that arbitrary
functions of time can be added, and in fact there is a more symmetrical form
\begin{equation}
  H_{S} = \half f_0 f_1 f_2
                + \third (\alpha_1-\alpha_2)f_0
                + \third (\alpha_1+2\alpha_2)f_1
                - \third (2\alpha_1+\alpha_2)f_2 \ ,
\end{equation}
which is central to the Okamoto theory (termed the auxiliary Hamiltonian).
However for our purposes this complicates some later results so we prefer the 
unsymmetrical form. Furthermore, in the full theory of PIV 
\cite{NY-97,NY-2000}
the Hamiltonian $H_0 \equiv H$ is associated with two additional
Hamiltonians $H_1 = \pi(H_0), H_2 = \pi^2(H_0) $ but these are not
required in the random matrix context.

It is also true that $ H(t) $ can be specified as the solution of a certain
second order second degree equation.
\begin{proposition}[\cite{Ok-86,JM-81}]
The Hamiltonian (\ref{7.a}) satisfies the second order second degree 
differential equation of the Jimbo-Miwa-Okamoto $ \sigma $ form for PIV,
\begin{equation}\label{f.5}
(H'')^2 - 4(tH' - H)^2 + 4 H'(H'+2 \alpha_1)(H' - 2\alpha_2) = 0.
\end{equation}
\end{proposition}
\noindent {\it Proof}.
Making use of Hamilton's equations (\ref{7.1}), we have for
$ H(t) = H(t;q(t),p(t)) $,
\begin{align}
   H'  & = f_1 f_2 \label{f.0} \ ,\\
   H'' & = f_1 f_2(f_2-f_1) + 2\alpha_2 f_1 + 2\alpha_1 f_2 . 
\label{f.1}
\end{align}
Use of (\ref{7.a}) and (\ref{f.0}) in (\ref{f.1}) shows
\begin{equation}
\begin{split}
f_1 & = {-\half H'' + t H' - H \over H' - 2\alpha_2} ,\\
f_2 & = { \half H'' + t H' - H \over H' + 2\alpha_1} .
\end{split}
\label{PIV-H2qp}
\end{equation}
Substituting (\ref{PIV-H2qp}) in (\ref{7.a}) gives the desired equation
(\ref{f.5}).
\qed

For future reference we note that use of Tables \ref{A2-param}, \ref{A2-Bxfm} 
shows that under the
action of the generators of $\widetilde{W}$, $H$ transforms according to
\begin{equation}
\begin{split}
   s_0(H) & = H + {2\alpha_0 \over f_0} \ ,\\
   s_1(H) & = H + 2 \alpha_1 t \ ,\\
   s_2(H) & = H - 2 \alpha_2 t \ ,\\
   \pi(H) & = H + f_2 - 2 \alpha_2 t.
\end{split}
\label{f0}
\end{equation}

\subsection{Toda lattice equation}
The $\tau$-function $\tau = \tau(t)$ is defined in terms of the Hamiltonian
$H(t)$ by
\begin{equation}\label{tau}
H(t) = {d \over dt} \log \tau(t).
\end{equation}
It is possible to derive a Toda lattice equation for the sequences of 
$\tau$-functions $\{\tau_k[n]\}_{n=0,1,\dots}$ ($k = 1,3$) associated
with the Hamiltonians
\begin{equation}\label{7.2}
H \Big |_{\alpha_0 \mapsto \alpha_0 + n \atop
\alpha_{1} \mapsto \alpha_{1} - n}, \qquad
H \Big |_{\alpha_0 \mapsto \alpha_0 + n \atop
\alpha_{2} \mapsto \alpha_{2} - n} \ ,
\end{equation}
respectively (the reason for the subscripts 1 and 3 on $ \tau $ will become 
apparent subsequently). An essential point is that there exist shift operators 
from the algebra $\widetilde{W}$ which after $n$ applications on $H$
generate the shifts required by (\ref{7.2}). There are in fact three
fundamental shift operators \cite{NY-97}
$ T_1 := \pi s_2 s_1, T_2 := s_1 \pi s_2, T_3 := s_2 s_1 \pi $ corresponding 
to translations on the root lattice by the fundamental weights 
$ \tilde{\omega}_j, \, j=1,2,3 $ of the root system $ A^{(1)}_2 $.
As can be checked from Tables \ref{A2-param}, \ref{A2-Bxfm} and (\ref{f0}) 
these operators have the property that
\begin{equation}\label{ap1}
T_1 H = H \Big |_{\alpha_0 \mapsto \alpha_0 + 1 \atop
\alpha_{1} \mapsto \alpha_{1} - 1}, \qquad
T^{-1}_3 H = H \Big |_{\alpha_0 \mapsto \alpha_0 + 1 \atop
\alpha_{2} \mapsto \alpha_{2} - 1} \ .
\end{equation}
Table \ref{A2-param} also shows that when acting on the parameters themselves, 
the same shifts occurring in the transformed Hamiltonian results, and thus
\begin{equation}
\begin{split}
   T_1(\alpha_0, \alpha_1,\alpha_2) 
	& = (\alpha_0 +1, \alpha_1 -1, \alpha_2) \ ,\\
   T_3^{-1}(\alpha_0, \alpha_1,\alpha_2) 
	& = (\alpha_0 +1, \alpha_1,  \alpha_2 -1) \ .
\end{split} 
\label{gb}
\end{equation} 
After a further $n$ iterations the equations (\ref{ap1}) can be written
in the form
\begin{equation}\label{26}
T_1^{n+1}H - T_1^n H = f_{(1)2}[n], \qquad
T_3^{-(n+1)} H - T_3^{-n} H = - f_{(3)1}[n] \ ,
\end{equation}
where the subscripts $(1)$ ($(3)$) refer to the system of Eqs.
(\ref{c1}) with the parameters replaced as in the first (second)
Hamiltonian (\ref{7.2}) and use has been made of (\ref{7.a}).

We remark that the two results of (\ref{ap1}) 
are inter-related. Thus consider the mapping $\omega$
defined by multiplication by $-1$ together with the replacements
\begin{equation}\label{rmt1}
(\alpha_0,\alpha_1,\alpha_2) \mapsto
(-\alpha_0, - \alpha_2, - \alpha_1), \qquad
(f_0, f_1, f_2) \mapsto (-f_0, - f_2, - f_1).
\end{equation}
We see immediately that the system (\ref{c1}) is unchanged by $\omega$, as is
the Hamiltonian (\ref{7.a}), while we can check from Table \ref{A2-param} that
\begin{equation}\label{rmt2}
\omega T_1 \omega = T_3^{-1}.
\end{equation}
Applying $\omega$ to the first equation of (\ref{ap1}) using
(\ref{rmt1})
and (\ref{rmt2}) gives the second equation. With the $\tau$-functions
$\tau_1[n]$ and $\tau_3[n]$ defined by
\begin{equation}\label{rmt3b}
T_1^n H = {d \over dt} \log \tau_1[n], \qquad
T_3^{-n} H = {d \over dt} \log \tau_3[n] \ ,
\end{equation}
application of (\ref{rmt2}) shows
\begin{equation}\label{rmt3}
\omega \tau_1[n] = C \tau_3[n] .
\end{equation}

In light of the relation (\ref{rmt3}), let us focus attention on the first 
equation of (\ref{ap1}) only. 
\begin{proposition}[\cite{Ok-86,KMNOY-99}]
The $ \tau$-function sequence $ \tau_1[n] $ corresponding to the parameter
sequence $ (\alpha_0+n, \alpha_1-n, \alpha_2) $ obeys the Toda lattice equation
\begin{equation}\label{tl}
{d^2 \over  dt^2} \log\sigma_1[n] =
	{\sigma_1[n\+ 1]\sigma_1[n\m 1] \over \sigma^2_1[n]}\ ,
\end{equation}
where
\begin{equation}\label{pu'}
\sigma_1[n] := C e^{t^2(\alpha_{1} - n)} \tau_1[n]\ .
\end{equation}
\end{proposition}
\noindent {\it Proof}.
Following \cite{Ok-86,KMNOY-99} we make use of the first equation in (\ref{26})
and consider the difference
\begin{equation}\label{7.ab}
\Big ( T_1^{n+1}  H -  T_1^n H \Big ) -
\Big (  T_1^n H - T_1^{n-1} H \Big ) =
f_{(1)2}[n] - T_{1}^{-1} f_{(1)2}[n] .
\end{equation}
A crucial fact, which follows from Table \ref{A2-Bxfm} and (\ref{7.1}), is 
that this difference is a total derivative
\begin{equation}\label{u1}
f_{(1)2}[n] - T_{1}^{-1} f_{(1)2}[n] = {d \over dt}
	\log \Big ( f_{(1)1}[n] f_{(1)2}[n] + 2(\alpha_{1} - n) \Big ).
\end{equation}
But it follows from (\ref{f.0}) and (\ref{7.a}) that 
\begin{equation}
 f_{(1)1}[n] f_{(1)2}[n] + 2(\alpha_{1} - n) =
    {d \over dt} \Big ( T_1^n H + 2 t(\alpha_{1} - n) \Big ) =
    {d^2 \over dt^2} \log \Big( e^{t^2 (\alpha_{1} - n)} \tau_1[n] \Big) \ ,
\end{equation}
and hence the right-hand side of (\ref{u1}) is equal to 
\begin{equation}\label{u3'}
{d \over dt} \log \left(
{d^2 \over dt^2} \log\left( e^{t^2 (\alpha_{1} - n)} \tau_1[n] \right)\right).
\end{equation}
On the other hand (\ref{7.ab}) and (\ref{rmt3b}) shows the left-hand side of
(\ref{u1}) is equal to
\begin{equation}\label{u3''}
{d \over dt} \log\left( {\tau_1[n\+ 1]\tau_1[n\m 1] \over \tau^2_1[n] }\right).
\end{equation}
Equating (\ref{u3'}) and (\ref{u3''}), and integrating shows that
\begin{equation}\label{pu}
{d^2 \over dt^2} \log \left( e^{t^2 (\alpha_{1} - n)} \tau_1[n] \right) = 
	C {\tau_1[n\+ 1]\tau_1[n\m 1] \over \tau^2_1[n] } ,
\end{equation}
and the stated result (\ref{tl}) follows. There is the ambiguity of a 
multiplicative constant $ C $, possibly dependent on $ n $ but not on $ t $,
and this can be chosen freely, for example to render the Toda lattice 
equation in a simple form.
\qed

The Toda lattice equation obeyed by the $\tau_3[n]$ with parameters 
$ (\alpha_0+n, \alpha_1, \alpha_2-n ) $ is obtained by applying the mapping 
$\omega$ to both sides of (\ref{pu}) and making use of (\ref{rmt3}). 
This shows
\begin{equation}\label{puk}
{d^2 \over dt^2} \log \left( e^{-t^2 (\alpha_{2} - n)} \tau_3[n] \right)
   = C{\tau_3[n\+ 1]\tau_3[n\m 1] \over \tau^2_3[n]} \ ,
\end{equation}
which with
\begin{equation}\label{pu'a}
\sigma_3[n] := C e^{-t^2(\alpha_{2} - n)} \tau_3[n] \ ,
\end{equation} 
gives the Toda lattice equation
\begin{equation}\label{tla}
{d^2 \over  dt^2} \log\sigma_3[n] =
	{\sigma_3[n\+ 1]\sigma_3[n\m 1] \over \sigma^2_3[n]}.
\end{equation}

Another way to deduce (\ref{puk}) is via (\ref{pu}) and the
differential equation (\ref{f.5}). Now, the second Hamiltonian in
(\ref{7.2}) is obtained from the first by simply interchanging
$\alpha_1$ and $\alpha_2$. On the other hand $\alpha_1$ and $\alpha_2$
are interchanged in (\ref{f.5}) if we replace $t$ by $it$ then replace
$H(it)$ by $-i H(t)$. This tells us that in (\ref{pu}) we can make
the replacements
\begin{equation}\label{rpl}
t \mapsto it, \quad \tau_1(it) \mapsto \tau_3(t), \quad \alpha_1
\mapsto \alpha_2,
\end{equation}
which indeed gives (\ref{puk}). Furthermore, since (\ref{rpl}) shows
$\tau_1$ and $\tau_3$ are simply related, it suffices to consider one
sequence only, $\tau_3[n]$ say.

\subsection{Classical solutions}
For a special initial choice of the parameters it is possible to choose
$\tau_3[0]=1$, and then to determine $\tau_3[1]$ in terms of a
classical function, that is to say the solution of a second order linear
differential equation. What is essential here is the condition for the
decoupling of the two independent first order differential equations so that
what remains is a Riccati equation.

\begin{proposition}[\cite{Ok-86}]
For the special initial choice of the parameters such that $ \alpha_2 = 0 $,
i.e. for parameters $(1-\alpha_1,\alpha_1,0)$ the first nontrivial member of
the $ \tau$-function sequence $ \tau_3[1] $ satisfies the Hermite-Weber 
equation, 
\begin{equation}\label{2.42a}
\tau_3''[1] = -2 t \tau_3'[1] -2 \alpha_1 \tau_3[1] .
\end{equation}
\end{proposition}
\noindent {\it Proof}.
With $n=0$, $ (1-\alpha_1+n,\alpha_1,-n) $ implies $\alpha_2=0$. Now we see 
from (\ref{7.a}) that
\begin{equation} 
H \Big |_{\alpha_2 =0} = pq(2p - q - 2t) -2 \alpha_1 p \ ,
\end{equation} 
which allows us to take 
\begin{equation}\label{ps3}
H \Big |_{\alpha_2 =0} = 0 \ ,
\end{equation}
provided we set $p=0$. Recalling (\ref{rmt3b}) this implies 
\begin{equation}\label{dis}
\tau_3[0] = 1 \ .
\end{equation}
We read off from the $n=1$ case of the second equation in (\ref{26}),
together with (\ref{12'}), that
\begin{equation}
T_3^{-1} H \Big |_{\alpha_2 =0} - H \Big |_{\alpha_2 =0} = q,
\end{equation}
and thus, after recalling the second equation in (\ref{rmt3b}) and (\ref{ps3}),
\begin{equation}\label{sb}
{d \over dt} \log \tau_3[1] = q.
\end{equation}
The first of Hamilton's equations (\ref{7.1}) gives, with 
$\alpha_2 =0$ and $p=0$ (after the differentiation), the Riccati equation
\begin{equation}
q' = -q^2 - 2tq -2 \alpha_1.
\end{equation}
Substituting (\ref{sb}) this reduces to the linear equation (\ref{2.42a})
first obtained in the present context by Okamoto \cite{Ok-86}.
\qed

\begin{proposition}
Two linearly independent solutions to the Toda lattice equation (\ref{puk})
for sequences of $ \tau$-functions with parameters 
$ (\alpha_0+n, \alpha_1, -n) $, $ n \geq 0 $, starting from the 
Weyl chamber wall $ \alpha_2 = 0 $ are given by the determinant forms
\begin{equation}\label{tu1}
\tau_3[n](t;\alpha_1)  = C
\det \Big [ \int_{-\infty}^t (t - x)^{-\alpha_1 + i+j} e^{-x^2} \, dx
\Big ]_{i,j=0,\dots,n-1} \ ,
\end{equation}
and
\begin{equation}\label{tu1'}
\bar{\tau}_3[n](t;\alpha_1)  = C
\det \Big [ \int_{-\infty}^\infty (t - x)^{-\alpha_1 + i+j} e^{-x^2} \, dx
\Big ]_{i,j=0,\dots,n-1} \ .
\end{equation}
\end{proposition}
\noindent {\it Proof}.
In the special case $\alpha_1 = 0$, we observe that a  solution of 
(\ref{2.42a}) is
\begin{equation}\label{dis1}
\tau_3[1] = C \int_{-\infty}^t e^{-x^2} \, dx.
\end{equation}
In fact it is possible to solve (\ref{2.42a}) in a form analogous to
(\ref{dis1}) for general $\alpha_1$. Thus consider the integral
\begin{equation}\label{2.44z}
I_{a}(t) := \int_{-\infty}^t (t - x)^{a} e^{-x^2} \, dx,
\end{equation}
and suppose temporarily that Re$(a) > -1$.
Simple manipulation gives
\begin{align}
   I_{a}(t) 
   & = t I_{a-1}(t) + \half \int_{-\infty}^t (t - x)^{a-1}
   {d \over dx} e^{-x^2} \, dx \nonumber \\
   & = t I_{a-1}(t) + {(a-1) \over 2} I_{a-2}(t).
\label{2.53a}
\end{align}
But
\begin{equation}\label{2.53b}
(a-1) I_{a-2}(t) = 
{1 \over a} {d^2 \over dt^2} I_{a}(t), \qquad
I_{a-1}(t) = {1 \over a} {d \over dt} I_{a}(t) \ .
\end{equation}
Thus we see that $I_{-\alpha_1}(t)$ satisfies (\ref{2.42a}) and this implies
\begin{equation}\label{2.44a}
\tau_3[1] = C \int_{-\infty}^t (t - x)^{-\alpha_1} e^{-x^2} \, dx \ ,
\end{equation}
where we require Re$(\alpha_1) < 1$.

Starting with (\ref{dis}) and (\ref{2.44a}), up to a multiplicative constant
the $\tau$-functions $\tau_3[n]$ $(n=2,3,\dots)$ are uniquely
specified by the Toda lattice equation (\ref{tla}). 
In fact it was known to Sylvester (\cite{matrix-Me} pp.~115-117) that the 
solution of (\ref{tla}) with initial condition $\sigma_3[0] = 1$
is the double Wronskian or Hankel determinant
\begin{equation}\label{11.b}
\sigma_3[n] =
\det \left[ {d^{i+j} \over dt^{i+j}} \sigma_3[1] \right]_{i,j=0,\dots,n-1}.
\end{equation}
Recalling (\ref{pu'a}) and (\ref{2.44a}) we therefore have 
\begin{equation}\label{an}
\tau_3[n] := \tau_3[n](t;\alpha_1) = C \det \left[ e^{-t^2}
{d^{i+j} \over dt^{i+j}} \Big ( e^{t^2} \int_{-\infty}^t (t - x)^{-\alpha_1}
	e^{-x^2} \, dx \Big ) \right]_{i,j=0,\dots,n-1}.
\end{equation}
Making use of (\ref{2.53a}) we can check that
\begin{equation}\label{an1}
{d^p \over dt^p} \Big ( e^{t^2}
\int_{-\infty}^t (t - x)^{-\alpha_1} e^{-x^2} \, dx \Big ) =
2^p  e^{t^2} \int_{-\infty}^t (t - x)^{-\alpha_1+p} e^{-x^2} \, dx \ ,
\end{equation}
so (\ref{an}) can also be written in the final form of (\ref{tu1}).

The second linearly dependent solution of (\ref{2.42a}) can also be written
in an integral form similar to (\ref{2.44z}). Thus we see the integral
\begin{equation}\label{2.44z'}
\bar{I}_{a}(t) := \int_{-\infty}^\infty
 (t - x)^{a} e^{-x^2} \, dx,
\end{equation}
satisfies the formulas (\ref{2.53a}) and (\ref{2.53b}), and thus satisfies
(\ref{2.42a}) with $a = - \alpha_1$. 
Hence in addition to (\ref{2.44a}) we have the solution
\begin{equation}\label{2.44j}
\bar{\tau}_3[1]
= C \int_{-\infty}^\infty (t - x)^{-\alpha_1} e^{-x^2} \, dx \ ,
\end{equation}
(note that for $\alpha_1$ not equal to a non-positive integer, this is 
well defined only if $t \not\in \R$). Proceeding as in the derivation of
(\ref{tu1}), we deduce from the Toda lattice equation (\ref{tla}), and
the initial values (\ref{dis}), (\ref{2.44j}), the sequence of
$\tau$-functions given by (\ref{tu1'}).
\qed

Let us now consider the sequence of $\tau$-functions $\bar{\tau}_1[n]$.
\begin{proposition}[\cite{Ok-86}]
The sequence of $\tau$-function solutions to the Toda lattice equation 
(\ref{pu}) $\bar{\tau}_1[n]$, $ n \geq 0 $, corresponding to the parameter
sequence $ (\alpha_0+n, -n, \alpha_2) $ starting from the line $ \alpha_1 = 0 $
has the determinantal form
\begin{equation}\label{tqa}
\bar{\tau}_1[n](t;-p)  = C
\det \Big [ H_{p + i + j}(t) \Big ]_{i,j=0,\dots,n-1}
\end{equation}
for $ -\alpha_2 = p \in \Z_{\ge 0} $.
\end{proposition}
\noindent {\it Proof}.
This sequence can be obtained from $\bar{\tau}_3[n]$ by the (inverse of) the 
mappings (\ref{rpl}). Replacing $t$ by $-it$ in (\ref{tu1}) does not lead to 
an integral of interest in random matrix applications, but doing the same in 
(\ref{tu1'}) gives
\begin{equation}\label{tq}
\bar{\tau}_1[n](t;\alpha_2)  = C
\det \Big [ \int_{-\infty}^\infty (t - i x)^{-\alpha_2 + i+j} e^{-x^2} \, dx
\Big ]_{i,j=0,\dots,n-1} \ ,
\end{equation}
which is of interest. We recall that for $\bar{\tau}_1^{(n)}$ the
parameters $(\alpha_0,\alpha_1,\alpha_2)$ in the corresponding
Hamiltonian are given by
\begin{equation}\label{2.40i}
(1 - \alpha_2 + n, -n, \alpha_2).
\end{equation}
For $p \in \Z_{\ge 0}$ we know
\begin{equation}\label{herm}
   \int_{-\infty}^\infty (t - i x)^p  e^{-x^2} \, dx = 
   \sqrt{\pi}2^{-p} H_p(t),
\end{equation}
and thus setting $\alpha_2 = -p$ equation (\ref{tq}) yields (\ref{tqa}).
\qed

Note that with $p = N$, $n=2$, this is precisely the final determinant
in (\ref{1.1}). 

\subsection{B\"acklund transformations and discrete Painlev\'e systems}
It has been known that some of the B\"acklund transformations of the 
PIV transcendent can be identified with discrete Painlev\'e 
equations \cite{BCH-95,GR-98},
although no systematic study has been undertaken for this class.
We will find that the difference equations for $ f_j[n], H[n], \tau[n] $ 
which are
generated by the B\"acklund transformations for the two shift operations
$ T^{-1}_3, T_1 $ are in fact manifestations of discrete Painlev\'e equations.

\begin{proposition}[\cite{GR-98}]
The B\"acklund transformations of the PIV system corresponding to the 
shift operator $ T^{-1}_3 $ generating the parameter sequence of
$ (\alpha_0+n, \alpha_1, -n) $ with $ n \in \mathbb{Z} $, 
$ 0 < \alpha_0,\alpha_1 < 1 $ and $ \alpha_0+\alpha_1=1 $ are 
second order difference equations of the first discrete Painlev\'e equation dPI, 
namely
\begin{equation}
  \chi_{k+1}+\chi_{k}+\chi_{k-1} = 2t +
  {k - (\half +\alpha_1) + (-1)^k(\half -\alpha_1) \over \chi_{k}}
  \qquad k \geq 1 \ ,
\label{adPI-T3}
\end{equation}
where 
\begin{equation}
  \chi_{2n\+ 1} = f_{(3)2}[n], \quad \chi_{2n\+ 2} = f_{(3)0}[n] 
  \qquad n \geq 0 \ .
\label{chi-adPI-T3}
\end{equation}
\end{proposition}
\noindent {\it Proof}.
The action of the shift operators on the $ f_j $ is expressible in a terminating
continued fraction, which for $ T_3 $ and its inverse takes the form
\begin{align}
  T_3(f_0) 
  & = f_1 - \cfrac{2\alpha_2}{f_2}
  \ ,\label{Bxfm-T3:a} \\
  T_3(f_1)
  & = f_2 + \cfrac{2(\alpha_1\+ \alpha_2)}{f_1 - \cfrac{2\alpha_2}{f_2}}
  \ ,\label{Bxfm-T3:b} \\
  T_3(f_2)
  & = f_0 + \cfrac{2\alpha_2}{f_2}
          - \cfrac{2(\alpha_1\+ \alpha_2)}{f_1 - \cfrac{2\alpha_2}{f_2}}
  \ ,\label{Bxfm-T3:c} \\
  T^{-1}_3(f_0)
  & = f_2 - \cfrac{2\alpha_0}{f_0}
          + \cfrac{2(\alpha_0\+ \alpha_1)}{f_1 + \cfrac{2\alpha_0}{f_0}}
  \ ,\label{Bxfm-T3:d} \\
  T^{-1}_3(f_1)
  & = f_0 - \cfrac{2(\alpha_0\+ \alpha_1)}{f_1 + \cfrac{2\alpha_0}{f_0}}
  \ ,\label{Bxfm-T3:e} \\
  T^{-1}_3(f_2)
  & = f_1 + \cfrac{2\alpha_0}{f_0}
  \ ,\label{Bxfm-T3:f}
\end{align}
as one can verify using the action of the affine Weyl group reflections and
diagram rotations as given in Tables \ref{A2-param}, \ref{A2-Bxfm}.
For simplicity of notation we suppress the subscript $(3)$ labelling the 
sequence $ (\alpha_0+n, \alpha_1, -n) $ during the discussion of our proofs as 
there is no risk of confusion.
Taking the first and last members of this set, now at the $ n^{\rm th} $ rung of the
$ T_3 $ ladder, and adding their unshifted $ f$-variable we have
\begin{equation}
  \begin{split}
  f_0[n] + f_0[n\m 1] & = 2t - f_2[n] + {2n \over f_2[n]}
  \ ,\\
  f_2[n\+ 1] + f_2[n] & = 2t - f_0[n] + {2(n\+ \alpha_0) \over f_0[n]}
  \ ,\end{split}
\label{fshift-T3}
\end{equation}
so that one has a closed system.
This can be recognised as the two components of a staggered system of 
difference equations and employing the definitions of $ \chi_k $ above we
arrive at the discrete Painlev\'e equation dPI.
\qed

In terms of the coordinate and momenta of the Hamiltonian system this 
difference system was found by Okamoto \cite{Ok-86,Ok-96} and can be expressed
as 
\begin{align}                                                        
    q[n\+ 1] 
  & = (2t\+ q[n]\m 2p[n]) {q[n](2t\+ q[n]\m 2p[n]) + 2\alpha_1 \over
         2(n\+ \alpha_0) - q[n](2t\+ q[n]\m 2p[n]) }
  \ ,\label{qpshift-T3:a}\\
    q[n\m 1]
  & = -2p[n] { q[n]p[n] - \alpha_1 \over q[n]p[n] - n}
  \ ,\label{qpshift-T3:b}\\
    p[n\+ 1] 
  & = - \half q[n] + { \alpha_0 + n \over 2t\+ q[n]\m 2p[n] }
  \ ,\label{qpshift-T3:c}\\
    p[n\m 1]
  & = t + { q[n]p[n] - n \over 2p[n] }
      - p[n]{ q[n]p[n] - \alpha_1 \over q[n]p[n] - n }
  \ .\label{qpshift-T3:d}
\end{align}
Consequently a third order difference equation exists for the Hamiltonian
through the relation
\begin{equation}
  H[n+1] - H[n] = -f_{1}[n] \ .
\label{Ham-adPI-T3}
\end{equation}
Eliminating $ p[n] $ between (\ref{qpshift-T3:a}), (\ref{qpshift-T3:b}) we 
find a second order difference equation for $ q[n] $,
\begin{equation}
\begin{split}
 & nq[n]q[n\m 1]\Big( 4t+2q[n]+q[n\+ 1]+q[n\m 1] \Big)^2
 \\
 &\quad
   = 2\Big[ (n\+1)q[n\+ 1]-nq[n\m 1]
	 - (2t+q[n]+q[n\+ 1])(\alpha_1 + \half q[n](2t+q[n])) \Big]
 \\
 & \qquad \times
      \Big[ (n\+1)q[n\+ 1]-nq[n\m 1] 
         - (2t+q[n]+q[n\m 1])(-\alpha_1 + \half q[n](2t+q[n]+q[n\+ 1]+q[n\m 1]))
      \Big] \ .
\end{split}
\label{qdiff-T3}
\end{equation}
Use of $ H[n\+ 1] - H[n] = q[n] $ leads to the third order equation in 
$ H[n] $.

In addition we have a higher order difference equation for the 
$ \tau$-function. 
\begin{proposition}
The $ \tau$-function sequence, appropriately normalised, associated with the 
shift operator $ T^{-1}_3 $ with parameter values 
$ (\alpha_0+n, \alpha_1, -n) $, $ n \in \Z_{\geq 0} $,
$ 0 < \alpha_0, \alpha_1 < 1 $ and $ \alpha_0+\alpha_1=1 $ satisfies the 
fourth order difference equation
\begin{equation}
\begin{split}
  & 4t^2\, \Big( 2n\tau^2[n]-\tau[n\+ 1]\tau[n\m 1] \Big) \times
           \Big( 2(n\m \alpha_1)\tau^2[n]-\tau[n\+ 1]\tau[n\m 1] \Big)
  \\
  & \quad\times
    \Big( \tau[n\m 2]\tau[n\+ 1]\tau[n] + 2\tau^2[n\m 1]\tau[n\+ 1]
		+ 4n(\alpha_1\m n)\tau^2[n]\tau[n\m 1] \Big)
  \\
  & \quad\times
    \Big( \tau[n\+ 2]\tau[n\m 1]\tau[n] - 2\tau^2[n\+ 1]\tau[n\m 1]
		+ 4n(\alpha_1\m n)\tau^2[n]\tau[n\+ 1] \Big)
  \\
  & \qquad = \Bigg\{
    \tau[n\+ 2]\tau[n\m 2]\tau^3[n] - 16n^2(\alpha_1\m n)^2\tau^5[n]
  \\
  & \qquad\quad
    + 16n(\alpha_1\m n)(\alpha_1\m 2n)\tau^3[n]\tau[n\+ 1]\tau[n\m 1]
    - 4(2n^2\m 2\alpha_1 n\+ 1)\tau[n]\tau^2[n\+ 1]\tau^2[n\m 1]
  \\
  & \qquad\qquad
    + \tau[n\+ 2]\tau^2[n\m 1]
      \Big( \tau[n\+ 1]\tau[n\m 1]+2(\alpha_1\+ 1\m 2n)\tau^2[n] \Big)
  \\
  & \qquad\qquad
    + \tau[n\m 2]\tau^2[n\+ 1]
      \Big( \tau[n\+ 1]\tau[n\m 1]+2(\alpha_1\m 1\m 2n)\tau^2[n] \Big)
    \Bigg\}^2 \ .
\end{split}
\label{taudiff-T3}
\end{equation}
\end{proposition}
\noindent {\it Proof}.
We first seek to express all the fundamental quantities in terms of the 
product $ f_1[n]f_2[n] $. By multiplying the two transformations 
(\ref{Bxfm-T3:a}) and (\ref{Bxfm-T3:b}) we find a quadratic relation for 
$ f_1[n] $ (and $ f_1[n\m 1] $),
\begin{equation}
   f_1[n](2t-f_1[n]) = f_1[n\+ 1]f_2[n\+ 1]+f_1[n]f_2[n]+2\alpha_1 \ .
\end{equation}
Next we multiply (\ref{Bxfm-T3:b}) by $ f_1[n] $ which yields a relation
for the product
\begin{equation}
   f_1[n]f_1[n\m 1] = f_1[n]f_2[n]
	{f_1[n]f_2[n]+2\alpha_1 \over f_1[n]f_2[n]+2n} \ .
\end{equation}
One can verify then that a linear proportionality exists between $ f_1[n] $
and $ f_1[n\m 1] $ via the product $ f_1[n]f_2[n] $,
\begin{equation}
\begin{split}
 & f_1[n] \left\{ f_1[n]f_2[n]+f_1[n\m 1]f_2[n\m 1]+2\alpha_1
	- f_1[n]f_2[n]{f_1[n]f_2[n]+2\alpha_1 \over f_1[n]f_2[n]+2n} \right\}
 \\ 
 & \quad =
   f_1[n\m 1] \left\{ f_1[n\+ 1]f_2[n\+ 1]+f_1[n]f_2[n]+2\alpha_1
        - f_1[n]f_2[n]{f_1[n]f_2[n]+2\alpha_1 \over f_1[n]f_2[n]+2n} \right\}
   \ ,
\end{split}
\end{equation}
so that $ f_1[n] $ and $ f_1[n\m 1] $ may now be linearly related to 
$ f_1[n]f_2[n] $. Multiplying these two later relations and using
\begin{equation}
  C {\tau_{3}[n\+ 1]\tau_{3}[n\m 1] \over \tau^2_{3}[n]}
  = 2n + f_{1}[n]f_{2}[n] \ ,
\label{tau-adPI-T3}
\end{equation}
with $ C = 1 $ to introduce the $ \tau$-functions we arrive at 
(\ref{taudiff-T3}).
\qed

\noindent {\it Note}.
The difference equations (\ref{qdiff-T3}) and (\ref{taudiff-T3}) have the
advantage of being of the lowest order we have found possible,
but the disadvantage of not being linear in the highest order
terms ($q[n+1]$ and $\tau[n+2]$ respectively). In fact
difference equations linear in the highest order terms can be
given by increasing by one the order of the equations in each
case \cite{Ok-96}.                                         

Applying the operator $\omega$ (recall (\ref{rmt1})) we obtain
analogous results for the sequence generated by $T_1$.
\begin{proposition}[\cite{GR-98}]
The B\"acklund transformations generated by the shift operator $ T_1 $ 
corresponding to the parameter sequence $ (\alpha_0+n, -n, \alpha_2) $
with $ n \in \mathbb{Z} $, $ 0 < \alpha_0, \alpha_2 < 1 $, and 
$ \alpha_0+\alpha_2 = 1 $, are second order difference equations of the 
first discrete Painlev\'e equation dPI, that is 
\begin{equation}
  \eta_{k+1} + \eta_{k} + \eta_{k-1} = 2t 
  - { k - [1\+ (-1)^k]\alpha_2 - \half[1\m (-1)^k] \over \eta_{k} } ,
  \qquad k \geq 1 \ ,
\label{adPI-T1}
\end{equation}
where
\begin{equation}
   \eta_{2n+1} = f_{(1)1}[n]\ , \qquad \eta_{2n+2} = f_{(1)0}[n]\ ,
   \qquad n \geq 0 \ .
\label{diag-R}
\end{equation}
\end{proposition}
\noindent {\it Proof}.
This follows immediately upon applying $\omega$ to both sides of 
(\ref{adPI-T3}).
\qed

The analogue of (\ref{qdiff-T3}) for the parameter sequence generated by the
shift $ T_1 $ can be found by applying the $\omega$ map to this relation, 
\begin{equation}
\begin{split}
 & -2np[n]p[n\m 1]\Big( 2t-2p[n]-p[n\+ 1]-p[n\m 1] \Big)^2
 \\
 &\quad
   =  \Big[ (n\+1)p[n\+ 1]-np[n\m 1]
         + (t-p[n]-p[n\+ 1])(\alpha_2 + 2p[n](t-p[n])) \Big]
 \\
 & \qquad \times
      \Big[ (n\+1)p[n\+ 1]-np[n\m 1]
         + (t-p[n]-p[n\m 1])(-\alpha_2 + 2p[n](t-p[n]-p[n\+ 1]-p[n\m 1]))
      \Big] \ ,
\end{split}
\label{pdiff-T1}
\end{equation}
and this implies a third order difference equation for the Hamiltonian via
\begin{equation}
   H[n\+ 1] - H[n] = f_{2}[n] = 2p[n] \ .
\label{Ham-adPI-T1}
\end{equation}
There is also a higher order difference equation for the $ \tau$-function
which can be derived using the relation
\begin{equation}
   {\tau_1[n\+ 1]\tau_1[n\m 1] \over \tau^2_1[n]} =
   f_{1}[n]f_{2}[n] - 2n \ ,
\label{tau-adPI-T1}
\end{equation}
although we do not reproduce this here.

\section{$ \tau$-Function Theory for PII}
\setcounter{equation}{0}
\subsection{Affine Weyl group symmetry}
In the general Painlev\'e theory the second Painlev\'e equation naturally 
appears as a coalescence limit of PIV. From the work of \cite{TW-94a} it is known 
that in random matrix theory PII occurs in the edge scaling limit of the GUE. 
This suggests that before studying this limit we should develop a theory of PII
analogous to that developed for PIV in the previous section.
We take the PII equation to be defined in the standard manner
\begin{equation}
   y'' = 2y^3 + ty + \alpha \ .
\label{PII}
\end{equation}

\begin{proposition}
The second Painlev\'e equation with the transcendent $ y=q(t) $ and parameter 
$ \alpha $ is equivalent to the system of first order differential equations
\begin{align}
   f_0' & = -2qf_0 + \alpha_0
   \ ,\label{aut-PII:a} \\
   f_1' & =  2qf_1 + \alpha_1
   \ ,\label{aut-PII:b}
\end{align}
where $ f_0+f_1 = 2q^2+t $ and $ \alpha_0 + \alpha_1 = 1 $ with 
$ \alpha = \alpha_1 - \half = \half - \alpha_0 $.
\end{proposition}
\noindent {\it Proof}.
This is established by eliminating $ p $ through the substitutions
$ f_0 = 2q^2 - p + t $ and $ f_1 = p $.
\qed
 
\begin{proposition}
Let $ \widetilde{W} = \langle s_0,s_1,\pi \rangle $ 
be the extended affine Weyl group of the root system of type $ A^{(1)}_1 $
generated by the reflections $ s_0, s_1 $ and the diagram rotation $ \pi $,
with action on the roots $ \alpha_0, \alpha_1 $
as given in Table \ref{A1-param},

\begin{table}[H]
\begin{center}
\begin{tabular}{|c|c|c|}
\hline

& $ \alpha_0 $ & $ \alpha_1 $ \\
\hline

  $ s_0 $ 
& $ -\alpha_0 $ 
& $ \alpha_1 + 2\alpha_0 $ \\
\hline

  $ s_1 $ 
& $ \alpha_0 + 2\alpha_1 $ 
& $ -\alpha_1 $ \\
\hline

  $ \pi $ 
& $ \alpha_1 $ 
& $ \alpha_0 $ \\
\hline

\end{tabular}
\caption{Action of the generators of the extended affine Weyl group associated
with the root system $ A^{(1)}_1 $ on the simple roots.}\label{A1-param}
\end{center}
\end{table}

The coupled system (\ref{aut-PII:a}), (\ref{aut-PII:b}) is symmetric under 
the B\"acklund transformations induced by the elements of the above 
affine Weyl group as specified in Table \ref{A1-Bxfm},

\begin{table}[H]
\begin{center}
\begin{tabular}{|c|c|c|c|}
\hline

& $ f_0 $ & $ f_1 $ & $ q $ \\
\hline

  $ s_0 $
& $ f_0 $
& \begin{minipage}[c]{3cm}{
  \begin{equation*}
     f_1 - {4\alpha_0 q\over f_0} + {2\alpha^2_0\over f^2_0}
  \end{equation*}
  }\end{minipage}
& \begin{minipage}[c]{2cm}{
  \begin{equation*}
     q - {\alpha_0 \over f_0}
  \end{equation*}
  }\end{minipage} \\
\hline

  $ s_1 $
& \begin{minipage}[c]{3cm}{
  \begin{equation*}
     f_0 + {4\alpha_1 q\over f_1} + {2\alpha^2_1\over f^2_1}
  \end{equation*}
  }\end{minipage}
& $ f_1 $
& \begin{minipage}[c]{2cm}{
  \begin{equation*}
     q + {\alpha_1 \over f_1}
  \end{equation*}
  }\end{minipage} \\
\hline

  $ \pi $
& $ f_1 $
& $ f_0 $
& $ -q $ \\
\hline

\end{tabular}
\caption{B{\"a}cklund transformations for the PII system.}\label{A1-Bxfm}
\end{center}
\end{table}

\end{proposition}
\noindent {\it Proof}.
This can be directly verified using the equations of motion 
(\ref{aut-PII:a}), (\ref{aut-PII:b}).
\qed

Underlying the dynamics of the PII system is a Hamiltonian structure.
\begin{proposition}
The PII dynamical system is equivalent to a Hamiltonian system $ \{q,p;H\} $
with Hamiltonian
\begin{equation}
  H = -\half f_0f_1 - \alpha_1 q \ ,
\label{Ham-PII}
\end{equation}
and canonical coordinates and momenta $ q, p $ defined by
\begin{equation}
     p    = f_1 \ , \qquad 2q^2 = f_0+f_1 - t \ .
\label{coord-PII}
\end{equation}
\end{proposition}
\noindent {\it Proof}.
Using the symmetrised differential equations (\ref{aut-PII:a},\ref{aut-PII:b})
the Hamilton equations of motion
\begin{equation}
  q' = p - q^2 - \half t\ , \qquad
  p' = 2qp + \alpha_1 \ ,
\label{PII-Heqn}
\end{equation}
can be verified.
\qed

\noindent {\it Remark}.
The fundamental domain or Weyl chamber for PII can be taken as the interval
$ \alpha \in (-\half,0] $ or $ \alpha \in [0,\half) $, and so there
exist identities relating the transcendents and related quantities at the
endpoints of these intervals. In particular, denoting the transcendent 
$ q(t,\alpha) $ and with $ \epsilon^2 = 1 $, $ t = -2^{1/3}s $ we have 
\cite{Gr-99}
\begin{equation}
\begin{split}
   -\epsilon\, 2^{1/3} q^2(s,0)
   & = {d \over dt}q(t,\half\epsilon) - \epsilon\, q^2(t,\half\epsilon) 
	- \half\epsilon\, t \ ,\\
   q(t,\half\epsilon) & =
   \epsilon\, 2^{-1/3} {1 \over q(s,0)} {d \over ds}q(s,0) \ .
\end{split}
\label{PII-ends}
\end{equation}

The action of the affine Weyl group on the Hamiltonian is given in
Table \ref{A1-Ham},

\begin{table}[H]
\begin{center}
\begin{tabular}{|c|c|c|}
\hline

& $ H_0 $ & $ H_1 = \pi(H_0) = H_0 + q $ \\
\hline

  $ s_0 $
& \begin{minipage}[c]{2cm}{
  \begin{equation*}
     H_0 + {\alpha_0 \over f_0}
  \end{equation*}
  }\end{minipage}
& $ H_1 $ \\
\hline

  $ s_1 $
& $ H_0 $
& \begin{minipage}[c]{2cm}{
  \begin{equation*}
     H_1 + {\alpha_1 \over f_1}
  \end{equation*}
  }\end{minipage} \\
\hline

  $ \pi $
& $ H_1 $
& $ H_0 $ \\
\hline

\end{tabular}
\caption{B{\"a}cklund transformations of the Hamiltonian.}\label{A1-Ham}
\end{center}
\end{table}

We define two shift operators corresponding to translations by the fundamental
weights of the affine Weyl group $ A^{(1)}_1 $,
\begin{equation}
   T_1 = \pi s_1, \qquad T_2 = s_1\pi \ ,
\label{shift-A1}
\end{equation}
although $ T_1 T_2 = 1 $ so only one is independent. Their action on the 
parameter space is given in Table \ref{A1-shift},

\begin{table}[H]
\begin{center}
\begin{tabular}{|c|c|c|}
\hline

& $ \alpha_0 $ & $ \alpha_1 $ \\
\hline

  $ T_1 $
& $ \alpha_0+1 $
& $ \alpha_1-1 $ \\
\hline

  $ T_2 $
& $ \alpha_0-1 $
& $ \alpha_1+1 $ \\
\hline

\end{tabular}
\caption{Action of the shift operators on the simple roots of the root system
$ A^{(1)}_1 $.}\label{A1-shift}
\end{center}
\end{table}

\begin{proposition}
The B\"acklund transformations corresponding to the shifts are given by
\begin{align}
 T_1(f_0) & = 
 f_1 - {4\alpha_0 q \over f_0} + {2\alpha^2_0 \over f^2_0}
 \ ,\label{Bxfm-shift-PII:a} \\
 T_1(f_1) & = f_0
 \ ,\label{Bxfm-shift-PII:b} \\
 T_2(f_0) & = f_1
 \ ,\label{Bxfm-shift-PII:c} \\
 T_2(f_1) & = 
 f_0 + {4\alpha_1 q \over f_1} + {2\alpha^2_1 \over f^2_1}
 \ ,\label{Bxfm-shift-PII:d} \\
 T_1(H_0) & = H_0 + q 
 \ ,\label{Bxfm-shift-PII:e} \\
 T_1(H_1) & = H_1 - q + {\alpha_0 \over f_0} 
 \ ,\label{Bxfm-shift-PII:f} \\
 T_2(H_0) & = H_0 + q + {\alpha_1 \over f_1}  
 \ ,\label{Bxfm-shift-PII:g} \\
 T_2(H_1) & = H_1 - q
 \ .\label{Bxfm-shift-PII:h}
\end{align}
\end{proposition}

\begin{proposition}
The Hamiltonian $ H(t) $ satisfies the second order second degree differential 
equation of Jimbo-Miwa-Okamoto $\sigma$ form for PII,
\begin{equation} 
  \left( H'' \right)^2 + 4\left( H' \right)^3 
  + 2H'[tH'-H] - \quarter\alpha^2_1 = 0 \ .
\label{PII-Hode}
\end{equation} 
\end{proposition}
\noindent {\it Proof}.
Using first two derivatives of $ H $ 
\begin{equation}
\begin{split}
  H'  & = -\half f_1  \ ,\\
  H'' & = -qf_1-\half\alpha_1 \ ,
\end{split}
\label{PII-derH}
\end{equation}
one can solve for $ q, f_0, f_1 $ and then substitute these back into the 
expression for $ H $. The result (\ref{PII-Hode}) then follows after 
simplification.
\qed

With $ H $ given by (\ref{Ham-PII}), and $ p $ and $ q $ specified by 
(\ref{coord-PII}), Hamilton's equation for $ q' $ implies
\begin{equation}
   p = q'+q^2+\half t \ .
\label{p-PII}
\end{equation}
Thus $ H $ can be expressed in terms of the Painlev\'e II transcendent $ q $ 
according to 
\begin{equation}
  H = \half (q')^2 - \half(q^2+\half t)^2 - (\alpha +\half)q \ .
\label{Hq-PII}
\end{equation}

\subsection{Toda lattice equation}
The $ \tau$-functions are defined as before (\ref{tau}) and corresponding to 
each sequence generated by the shift operators is a Toda lattice equation.
\begin{proposition}
The $ \tau$-function sequence generated by the shift operator $ T_1 $ with the
parameter sequence $ (\alpha_0+n, \alpha_1-n) $ for $ n \geq 0 $ 
\begin{equation}
   T^n_1(H) = H[n] = {d \over dt}\ln\tau[n] \ ,
\label{tau-PII}
\end{equation}
obeys the Toda lattice equation
\begin{equation}
  C {\tau[n\+ 1]\tau[n\m 1] \over \tau^2[n]} =
    {d^2 \over dt^2}\ln \tau[n] \ .
\label{Toda-PII}
\end{equation}
\end{proposition}
\noindent {\it Proof}.
This parallels the argument employed for PIV case, by utilising the relations
\begin{equation}
   H[n\+ 1] - H[n] = q[n] \ ,
\label{}
\end{equation}
and
\begin{equation}
  q[n] - T^{-1}_1(q[n]) = {d \over dt}\ln f_1[n] \ ,
\label{}
\end{equation}
along with
\begin{equation}
  {d \over dt}H = -\half f_1 \ .
\label{PII-dH}
\end{equation}
\qed

\subsection{Classical solutions}
When the parameter values are those on a chamber wall (a point) 
$ \alpha_1 = n \in \Z $ then the $ \tau$-functions are known to be expressible
in terms of Airy functions \cite{Ok-86}.
\begin{proposition}
The solution for the first non-trivial member of the $ \tau$-function sequence
$ \tau[1](t) $ generated by the shift operator $ T_1 $ with initial parameters
$ (\alpha_0, \alpha_1) = (1, 0) $ that is bounded as $ t \to -\infty $ is
\begin{equation}
  \tau[1](t) = C {\rm Ai}(-2^{-1/3}t) \ .
\label{PII-airy1}
\end{equation}
The $ n^{\text{th}}$ member of this sequence is 
\begin{equation}
  \tau[n](t) = C \det\left[ {d^{i+j} \over dt^{i+j}}{\rm Ai}(-2^{-1/3}t)
                     \right]_{i,j = 0,\ldots ,n-1} \ .
\label{Airy-det}
\end{equation}
\end{proposition}
\noindent {\it Proof}.
Starting from $ \alpha_1 = 0 $ at $ n=0 $ one can take
$ p[0] = 0 $ so that $ H[0] = 0 $ and conventionally $ \tau[0] = 1 $.
Using (\ref{Bxfm-shift-PII:e}) we find that $ H[1] = q[0] $ and so the equation 
of motion (\ref{PII-Heqn}) gives the second order linear differential equation
\begin{equation}
  \tau''[1] + \half t\tau[1] = 0 \ ,
\end{equation}
and thus (\ref{PII-airy1}).
The determinant formula (\ref{Airy-det}) follows from (\ref{11.b}).
\qed

Another special parameter value of the Hamiltonian system (\ref{Ham-PII}) is
$ \alpha_1 = -\half $ \cite{Vo-65} when Hamilton's equation (\ref{PII-Heqn})
permit the solution
\begin{equation}
  (q,p) = (t^{-1}, \half t) \ .
\label{PII-rational}
\end{equation}
However the corresponding value of $ H $ is not zero so in this case we do not
have $ \tau[0] = 1 $ (rather $ H[0] $ and thus $ \log\tau[0] $ is a rational
function of $ t $), and thus the sequence of $ \tau$-functions generated by
$ T_1 $ is not given by a determinant \cite{KM-99}. Nonetheless the B\"acklund
transformations of Prop. 15 show that $ H[n] $ and thus $ \log\tau[n] $
remain rational functions of $ t $ for all $ n = 1,2,3,\ldots $.

\subsection{B{\"a}cklund transformations and discrete dPI}
The discrete dynamical system generated by the B\"acklund transformations is
also integrable and can be identified with a discrete Painlev\'e system
\begin{proposition}
The members of the sequence $ \{q[n]\} $, $ n \geq 0 $
generated by the shift operator $ T_1 $
with the parameters $ (\alpha_0+n,\alpha_1-n) $
are related by a second order difference equation which is the alternate form 
of the first discrete Painlev\'e equation, a-dPI,
\begin{equation}
   {\alpha + \half - n \over q[n]+q[n\m 1]} +
   {\alpha - \half - n \over q[n\+ 1]+q[n]} =
   -2q^2[n] - t \ .
\label{adPI-PII}
\end{equation}
\end{proposition}
\noindent {\it Proof}.
We deduce from (\ref{shift-A1}) and Table \ref{A1-Bxfm} that
\begin{align}
   T_1(q) & = -q + {\alpha-\half \over p-2q^2-t}
   \ ,\label{} \\
   T_2(q) & = -q - {\alpha+\half \over p}
   \ ,\label{shift-q-PII}
\end{align}
so eliminating $ p $ through the combination of these two we arrive at the
stated result.
\qed

The full set of forward and backward difference equations are \cite{Ok-86}
\begin{align}
   q[n\+ 1] & = -q[n] + {\alpha-\half-n \over p[n]-2q[n]^2-t}
   \ ,\label{PII-Bxfm:a} \\
   q[n\m 1] & = -q[n] - {\alpha+\half-n \over p[n]}
   \ ,\label{PII-Bxfm:b} \\
   p[n\+ 1] & = -p[n]+2q[n]^2+t
   \ ,\label{PII-Bxfm:c} \\
   p[n\m 1] & = t - p[n] + 2\left( q[n] + {\alpha+\half-n \over p[n]} \right)^2
   \ .\label{PII-Bxfm:d}
\end{align}

The discrete Painlev\'e equation (\ref{adPI-PII}) implies a third order 
difference equation for the Hamiltonian
\begin{equation}
   {\alpha + \half - n \over H[n\+ 1]-H[n\m 1]} +
   {\alpha - \half - n \over H[n\+ 2]-H[n]} =
   -2(H[n\+ 1]-H[n])^2 - t \ ,
\label{HdiffE-PII}
\end{equation}
because $ q[n] = H[n+1]-H[n] $.
Equations (\ref{PII-Bxfm:c}) and (\ref{PII-Bxfm:d}) also imply
\begin{equation}\label{apn}
q[n] = {1 \over 4 \alpha_n} p[n] \left( p[n\m 1] - p[n\+ 1] \right)
     - {\alpha_n \over 2 p[n]},
 \quad \alpha_n := \alpha +\half -n.
\end{equation}
Using this to eliminate $ q[n] $ (in say (\ref{PII-Bxfm:c})) yields a second order
difference equation for the $p[n]$,
\begin{equation}\label{bh}
   {1 \over 4\alpha^2_n}p^2[n] \left( p[n\+ 1]-p[n\m 1] \right)^2
   + {\alpha^2_n \over p^2[n]} - 2p[n]-p[n\+ 1]-p[n\m 1] + 2t = 0 .
\end{equation}
Furthermore, Eqs. (\ref{tau-PII}), (\ref{Toda-PII}) and (\ref{PII-dH}) 
give
\begin{equation}
  C {\tau[n\+ 1] \tau[n\m 1] \over \tau^2[n]} = p[n].
\label{tau-p-PII}
\end{equation}
So substituting in (\ref{bh}) (with say $ C = -2 $) implies a fourth order 
difference equation for $ \tau[n] $,
\begin{equation}\label{rn2}
\begin{split}
 & {1 \over 2\alpha^2_n}
   \Big( \tau[n\m 2]\tau^2[n\+ 1] - \tau[n\+ 2]\tau^2[n\m 1]
   \Big)^2 + \eight\alpha^2_n\tau^6[n]
 \\
 & + \tau[n\m 2]\tau^3[n]\tau^2[n\+ 1]
   + \tau[n\+ 2]\tau^3[n]\tau^2[n\m 1]
   + 2\tau^3[n\m 1]\tau^3[n\+ 1]
   + t\;\tau^2[n]\tau^2[n\m 1]\tau^2[n\+ 1] = 0 \ .
\end{split}
\end{equation}
While (\ref{bh}) and the corresponding equation for $ \tau[n] $ provide a 
polynomial relation between the smallest set of consecutive sequence members
$ \{p[n]\} $ and $ \{\tau[n]\} $ we have found possible, they have the 
disadvantage of not being linear in the highest order term ($ p[n\+ 1] $ and
$ \tau[n\+ 2] $ respectively). This disadvantage can be remedied by increasing
by one the order of the equation in each case. Thus by replacing $ n $ by $ n-1 $
in (\ref{apn}), then adding the result to the original equation and using 
(\ref{PII-Bxfm:b}) implies a third order difference equation for the $ p[n] $
\cite{Ok-99},
\begin{equation}
 0 = {1 \over 4\alpha_n}p[n]\left(p[n\m 1]-p[n\+ 1]\right)
       + {\alpha_n \over 2p[n]}
       + {1 \over 4\alpha_{n-1}}p[n\m 1]\left(p[n\m 2]-p[n]\right) 
       - {\alpha_{n-1} \over 2p[n]} ,
\label{}
\end{equation}
which is indeed linear in $ p[n\+ 1] $. Substituting (\ref{tau-p-PII}) gives 
a fifth order equation for $ \tau[n] $, linear in the highest order term
$ \tau[n\+ 2] $.

\subsection{Coalescence from PIV}
Since the earliest works on the Painlev\'e transcendents \cite{Pa-1906} it 
was known how to obtain the PII system from a limiting procedure or 
coalescence applied to the PIV, however there is in fact more than one such 
coalescence path. In fact our analysis of the scaled GUE requires the 
application of a second coalescence path rather than the one commonly employed.
In the first limit the parameters $ (\alpha_0, \alpha_1, \alpha_2) $ and 
variables $ t_{\rm IV}, q_{\rm IV}, p_{\rm IV}, H_{\rm IV} $ in the PIV system
scale in a way that $ \alpha_2 $ is fixed so that
\begin{align}
   \alpha_0 & = \half - \alpha_{\rm II} - \half \epsilon^{-6} \ ,
            \label{PII-coal-1:a} \\
   \alpha_1 & = \half \epsilon^{-6} \ ,
            \label{PII-coal-1:b} \\
   \alpha_2 & = \alpha_{\rm II} + \half \ ,
            \label{PII-coal-1:c} \\
   t_{\rm IV} & = -\epsilon^{-3} + 2^{-2/3}\epsilon\, t_{\rm II} \ ,
            \label{PII-coal-1:d} \\
   q_{\rm IV} & =  \epsilon^{-3} + 2^{2/3}\epsilon^{-1} q_{\rm II} \ ,
            \label{PII-coal-1:e} \\
   p_{\rm IV} & =  2^{-2/3}\epsilon\, p_{\rm II} \ ,
            \label{PII-coal-1:f} \\
   H_{\rm IV} & = -\epsilon^{-3}(\alpha_{\rm II} + \half)
                  + 2^{2/3}\epsilon^{-1} H_{\rm II} \ ,
            \label{PII-coal-1:g}
\end{align}
as $ \epsilon \to 0 $ then the function $ q_{\rm II}(t_{\rm II}) $ satisfies
the PII differential equation with parameter $ \alpha_{\rm II} $
\cite{IKSY-91}.

The second limiting procedure is obtained from the first by the mapping 
$\omega$ introduced in (\ref{rmt1}).
\begin{proposition}
If $ \alpha_1 $ is fixed so that the variables scale like
\begin{align}
   \alpha_0 & = \alpha_{\rm II} + \thalf + \half \epsilon^{-6} \ ,
            \label{PII-coal-2:a} \\
   \alpha_1 & = -\alpha_{\rm II} - \half \ ,
            \label{PII-coal-2:b} \\
   \alpha_2 & = - \half \epsilon^{-6} \ ,
            \label{PII-coal-2:c} \\
   t_{\rm IV} & =  \epsilon^{-3} - 2^{-2/3}\epsilon\, t_{\rm II} \ ,
            \label{PII-coal-2:d} \\
   q_{\rm IV} & = 2^{1/3}\epsilon\, p_{\rm II} \ ,
            \label{PII-coal-2:e} \\
  2p_{\rm IV} & =  \epsilon^{-3} + 2^{2/3}\epsilon^{-1} q_{\rm II} \ ,
            \label{PII-coal-2:f} \\
   H_{\rm IV} + \alpha_1 t_{\rm IV} & = -2^{2/3}\epsilon^{-1} H_{\rm II} \ ,
            \label{PII-coal-2:g}
\end{align}
as $ \epsilon \to 0 $ then the function $q_{\rm II}(t_{\rm II})$ satisfies 
the PII differential equation with parameter $\alpha_{\rm II}$.
Furthermore the third order difference equation for $ H_{\rm IV} $ 
(\ref{pdiff-T1},\ref{Ham-adPI-T1}), related to the discrete Painlev\'e equation 
dPI, corresponding to the B\"acklund transformation under the shift operator 
$ T_1 $ transforms into the third order difference equation for $ H_{\rm II} $ 
(\ref{HdiffE-PII}), related to the alternate discrete Painlev\'e equation 
a-dPI, under this scaling.
\end{proposition}
\noindent {\it Proof}.
Under the mapping $ H_{\rm IV} \mapsto - H_{\rm IV} $, 
$ 2p_{IV} \leftrightarrow q_{IV} $, $ \alpha_1 \leftrightarrow -\alpha_2 $ 
and $ t_{IV} \leftrightarrow -t_{IV} $.
The only equation which isn't immediate from the mapping is 
(\ref{PII-coal-2:g}). Substituting (\ref{PII-coal-2:d}) for $ t_{\rm IV} $ and 
ignoring the term proportional to $ \epsilon\, t_{\rm II} $ shows this equation
is equivalent to
\begin{equation}
   H_{\rm IV} = \epsilon^{-3}(\alpha_{\rm II} + \half) 
                - 2^{2/3}\epsilon^{-1} H_{\rm II} ,
\end{equation}
which is precisely what results from applying the mapping $ \omega $ to
(\ref{PII-coal-1:g}). The scaling of the third order difference equation for 
$ H_{\rm IV} $ associated with the discrete Painlev\'e equation, dPI, 
(\ref{pdiff-T1}) to (\ref{HdiffE-PII}) can be verified directly.
\qed

\section{Application to Finite GUE Matrices}
\setcounter{equation}{0}
In this section we will show that the determinants in (\ref{tu1}) and
(\ref{tu1'}) occur in the calculation of the quantities $ \tilde{E}_N $ and
$ F_N $, introduced in the Introduction, relating to GUE random matrices.

\subsection{Calculation of $ E_N(0;(s,\infty)) $ and $ \tilde{E}_N(s;a) $}
Consider first the probability $E_N(0;(s,\infty))$. 
\begin{proposition}
The gap probability $ E_N(0;(s,\infty)) $ is identical to the $ N^{\text{th}}$ 
$ \tau$-function of the sequence generated by $ T^{-1}_3 $ from the corner of
the Weyl chamber $ (\alpha_0, \alpha_1, \alpha_2) = (1,0,0) $,
\begin{equation}\label{ni1}
E_N(0;(s,\infty)) = \tau_3[N](s;0) \ ,
\end{equation}
where the normalization of $\tau_3[N]$ must be such that
\begin{equation}\label{ni1'}
\lim_{s \to \infty} \tau_3[N](s;0) = 1.
\end{equation}
The resolvent kernel function $ R = R_N(t) $ occurring in (\ref{2}) is the 
$ N^{\text{th}}$ Hamiltonian associated with this sequence,
\begin{equation}
R_N(t) = H(t) \Big |_{(\alpha_0,\alpha_1,\alpha_2) = (1+N,0,-N)} \ .
\label{R-Ham}
\end{equation}
\end{proposition}
\noindent {\it Proof}.
From the meaning of $E_N$ we see from (\ref{1f}) with $g(x) = e^{-x^2}$ that
\begin{equation}\label{3.0}
E_N(0;(s,\infty)) = {1 \over C}
\int_{-\infty}^s dx_1 \cdots \int_{-\infty}^s dx_N \,
\prod_{j=1}^N e^{-x_j^2} \prod_{1 \le j < k \le N}
(x_k - x_j)^2 \ .
\end{equation}
Introducing the Vandermonde determinant
\begin{equation}\label{van}
\prod_{1 \le j < k \le N}
(x_k - x_j) = \det [x_{j+1}^k]_{j,k=0,\dots,N-1} \ ,
\end{equation}
standard manipulations following Heine (see \cite{ops-Sz} p.~27)
allow this to be rewritten as the $N \times N$ determinant
\begin{align}
   E_N(0;(s,\infty)) 
   & = {1 \over C} \det
     \Big[ \int_{-\infty}^s x^{j+k} e^{-x^2} \, dx \Big]_{j,k=0,\dots,N-1}
   \ ,\nonumber \\
   & = {1 \over C} \det
     \Big[ \int_{-\infty}^s (s - x)^{j+k} e^{-x^2} \, dx 
     \Big]_{j,k=0,\dots,N-1} \ ,
\label{aw}
\end{align}
where the second
equality follows by 
repeating the procedure which led to the first equality but starting from 
the Vandermonde determinant formula with 
$x_j \mapsto x_j - s$ $(j=1,\dots,N)$.
Comparing with (\ref{tu1}) gives (\ref{ni1}).
Recalling (\ref{rmt3b}) it follows from (\ref{ni1}) that
\begin{equation}\label{ni2}
   E_N(0;(s,\infty)) = \exp\Big( -\int_s^\infty 
   H(t) \Big |_{(\alpha_0,\alpha_1,\alpha_2) = (1+N,0,-N)}\; dt \Big) \ ,
\end{equation}
and comparing with (\ref{2}) gives (\ref{R-Ham}).
\qed

According to (\ref{f.5}) $R$ satisfies Eq. (\ref{c1a}), 
thus rederiving the result of Tracy and Widom (\ref{2}). 
We remark that the boundary condition follows from the fact that, with
$\rho(t)$ denoting the eigenvalue density,
\begin{equation}\label{denc}
R(t) \mathop{\sim}\limits_{t \to \infty} \rho(t) = 
	{2^{-N} \over \pi^{1/2} (N\m 1)!} e^{-t^2}
	\Big[ H^2_N(t) - H_{N+1}(t) H_{N-1}(t) \Big].
\end{equation}

Considering next $\tilde{E}_N(s;a)$ we have a generalisation of the previous
case.
\begin{proposition}
The average $\tilde{E}_N(s;a)$ is the $ N^{\text{th}}$ member of the 
$ \tau$-function sequence $ \{\tau_3[N](s;-a)\} $ generated by $ T^{-1}_3 $ 
from a general position on the $ \alpha_2 = 0 $ wall of the Weyl chamber, 
namely $ (\alpha_0, \alpha_1, \alpha_2) = (1+a,-a,0) $, and so
\begin{equation}\label{ent3}
   \tilde{E}_N(s;a) = \tau_3[N](s;-a).
\end{equation}
The logarithmic derivative $ U_N(t;a) $ is identical to the Hamiltonian
associated with this sequence,
\begin{equation}\label{UH}
   U_N(t;a) = H(t) \Big |_{(\alpha_0, \alpha_1, \alpha_2) = (1+a+N, -a, -N)}.
\end{equation}
\end{proposition}
\noindent {\it Proof}.
According to (\ref{1.3'}) $\tilde{E}_N(s;a)$ is given in terms of a multiple 
integral by
\begin{equation}\label{ent}
   \tilde{E}_N(s;a) = {1 \over C}
   \int_{-\infty}^s dx_1 \cdots \int_{-\infty}^s dx_N \,
   \prod_{j=1}^N e^{-x_j^2} (s - x_j)^a \prod_{1 \le j < k \le N}
   (x_k - x_j)^2,
\end{equation}
where the normalization $C$ is such that
\begin{equation}\label{ent1}
   \tilde{E}_N(s;a) \sim s^{Na} \Big ( 1 + O(1/s) \Big ) \tilde{E}_N(s;0)
   \quad {\rm as} \quad s \to \infty.
\end{equation}
The method of derivation of (\ref{aw}) shows that
\begin{equation}\label{ent2}
   \tilde{E}_N(s;a) = {1 \over C}
   \det \Big [ \int_{-\infty}^s (s - x)^{a+j+k} e^{-x^2} \, dx
   \Big ]_{j,k=0,\dots,N-1}.
\end{equation}
Recalling (\ref{tu1}) we thus have (\ref{ent3}). It follows from this that
\begin{equation}\label{ent4}
   \tilde{E}_N(s;a) = \tilde{E}_N(s_0;a) \exp \Big( \int_{s_0}^s 
    H(t) \Big|_{(\alpha_0, \alpha_1, \alpha_2) = (1+a+N, -a, -N)}
                                        \; dt \Big) \ ,
\end{equation}
and consequently (\ref{UH}).
\qed

According to (\ref{f.5}) $U_N(t;a)$ satisfies the nonlinear equation
\begin{equation}\label{rsw}
   (U_N'')^2 - 4(tU_N' - U_N)^2 + 4U_N'(U_N' - 2a)(U_N'+2N) = 0,
\end{equation}
which considering (\ref{ent1}) and (\ref{2}) is to be solved subject to 
the boundary condition
\begin{equation}\label{rsw1}
   U_N(t;a) \mathop{\sim}\limits_{t \to
\infty} {Na \over t} \Big ( 1 + O(1/t) \Big ) + R(t). 
\end{equation}
In light of (\ref{denc}) the term $R(t)$ decays as a Gaussian and so is
negligible with respect to the inverse powers in (\ref{rsw1}) (here we 
assume $a \ne 0$). This means
the boundary condition (\ref{rsw1}) cannot be distinguished from the
boundary condition with $R(t)$ removed; however we will see below that this
latter boundary condition relates to a solution of (\ref{rsw}) distinct
from the one required in (\ref{ent4}). To overcome this ambiguity we
specify the $t \to -\infty$ behaviour of $U(t)$ rather than the 
$t \to \infty$ behaviour. Now, replacing $s$ by $-s$ in (\ref{ent}) and
changing variables $x \mapsto x + s$ shows
\begin{align}
\tilde{E}_N(-s;a)
   & = {1 \over C} e^{-N s^2} s^{-Na - N^2}
	\int_0^\infty dx_1 \cdots \int_0^\infty dx_N \,
	\prod_{j=1}^N e^{-(x_j / s)^2} e^{-2x_j} x^a_j 
        \prod_{1 \le j < k \le N} (x_k - x_j)^2
   \nonumber \\
   & \sim {1 \over C} e^{-N s^2} s^{-Na - N^2} \Big( 1 + O(1/s^2) \Big).
\label{50}
\end{align}
In combination with (\ref{ent4}) this implies
\begin{equation}\label{rsw6}
   U_N(t;a) \mathop{\sim}\limits_{t \to -\infty}
   - 2N t - {N(a+N) \over t}  + O\Big ( {1 \over t^3} \Big ).
\end{equation}

Under the action of the B\"acklund transformations $U_N(t;a)$ will satisfy two
recurrence relations corresponding to the shift operators $ T^{-1}_3 $ and
$ T_1 $. For the $ T^{-1}_3 $ sequence we relate the Painlev\'e and GUE
parameters by $ n = N $ and $ \alpha_1 = -a $, and with (\ref{qdiff-T3}) and 
(\ref{Ham-adPI-T3}) we have a third order recurrence relation in $ N $ with
$ a $ fixed (suppressing the additional variable dependence in $ U $)
\begin{equation}
\begin{split}
 & -N(U_{N+1}-U_N)(U_N-U_{N-1})\Big( 4t+ U_{N+2}+U_{N+1}-U_N-U_{N-1} \Big)^2
 \\
 + 2 & \Big[ (N\+1)(U_{N+2}-U_{N+1})-N(U_N-U_{N-1})
 \\
 & \hskip2cm
         - \half(2t+U_{N+2}-U_N)(-2a+(U_{N+1}-U_N)(2t+U_{N+1}-U_N)) \Big]
 \\
 \times & \Big[ (N\+1)(U_{N+2}-U_{N+1})-N(U_N-U_{N-1})
 \\
 & \hskip2cm
         - \half(2t+U_{N+1}-U_{N-1})
                (2a+(U_{N+1}-U_N)(2t+U_{N+2}-U_{N-1})) \Big] = 0\ .
\end{split}
\label{Udiff-T3}
\end{equation}
In contrast the $ T_1 $ sequence has the parameter correspondence $ n = a $
and $ \alpha_2 = -N $ which is an interchange of $ \alpha_1 $ and $ \alpha_2 $
with respect to the $ T_3 $ sequence. A third order recurrence relation in 
$ a $, with fixed $ N $, can be derived from (\ref{Udiff-T3}) using the 
mapping (\ref{rpl}) in which $ N \leftrightarrow a $, $ t \mapsto it $ and then
$ U_N(it;a) \mapsto -iU_N(t;a) $.
Alternatively this difference equation can be found directly from 
(\ref{pdiff-T1}) and (\ref{Ham-adPI-T1}), with the identification (\ref{UH}).
% \begin{equation}
% \begin{split}
%  & a[U(a\+ 1)-U(a)][U(a)-U(a\m 1)]
%                \Big[ 4t-U(a\+ 2)-U(a\+ 1)+U(a)+U(a\m 1) \Big]^2
%  \\
%  + & 2\Big[ (a\+1)[U(a\+2)-U(a\+1)]-a[U(a)-U(a\m 1)]
%  \\
%  & \hskip2cm
%  + \half[2t+U(a)-U(a\+2)][-2N + [U(a\+ 1)-U(a)][2t+U(a)-U(a\+1))] \Big]
%  \\
%  \times & \Big[ (a\+1)[U(a\+2)-U(a\+1)]-a[U(a)-U(a\m 1)]
%  \\
%  & \hskip2cm
%  + \half[2t+U(a\m 1)-U(a\+1)][2N + [U(a\+ 1)-U(a)][2t+U(a\m 1)-U(a\+ 2)]
%    \Big]  = 0\ .
% \end{split}
% \label{Udiff-T1}
% \end{equation}

As noted in the Introduction, for $(N\+ 1) \times (N\+ 1)$ dimensional GUE
matrices, the distribution of the largest eigenvalue $p_{\rm max}(s)$ is
proportional to $e^{-s^2} \tilde{E}_N(s;2)$. Hence, using (\ref{ent4}),
\begin{equation}\label{ag}
p_{\rm max}(s) \Big |_{N \mapsto N +1} =
p_{\rm max}(s_0) \Big |_{N \mapsto N +1}
 \exp \Big( \int_{s_0}^s\left[ -2t + U_N(t;2) \right] \; dt \Big).
\end{equation}
Comparison with (\ref{1.1'}), after putting $N \mapsto N + 1$ therein
and substituting (\ref{2}), gives an identity between the Hamiltonians
$ R(t) \Big |_{N \mapsto N +1}$ and $U(t;2) $. Using the theory developed 
here this relation can be independently verified.
\begin{proposition}\label{pUR}
The logarithmic derivative of the average $\tilde{E}_N(t;2)$ is related to
the resolvent kernel $ R_N(t) $ by the identity
\begin{equation}
   U_N(t;2) = 2t + R_{N+1}(t) + {R'_{N+1}(t) \over R_{N+1}(t)} \ .
\end{equation}
\end{proposition}
\noindent {\it Proof}.
From the identifications made above we have
\begin{align}
   U_{N}(t;2) 
  & = H(t)\Big|_{(N+3,-2,-N)} \ ,\\
   R_{N+1}(t) 
  & = H(t)\Big|_{(N+2,0,-N-1)} \ ,
\end{align}
and we seek to express the first in terms of the second. We note that
\begin{align}
  H(t)\Big|_{(N+3,-2,-N)}
 & = T_1T^{-1}_2  H(t)\Big|_{(N+2,0,-N-1)} \ ,\\
 & = 2t + H + \cfrac{2(N+2)}{f_0}
        - \cfrac{2(N+1)}{f_0 + \cfrac{2}{f_2 - \cfrac{2(N+2)}{f_0}}} \ ,
\end{align}
where the right-hand side is evaluated at $ (N+2,0,-N-1) $.
We recognise in this expression the factors of
$ H|_{(N+2,0,-N-1)} = f_1(\half f_0f_2 -N -1) $ and
$ H'|_{(N+2,0,-N-1)} = f_1 f_2 $ so that it can be simplified to
\begin{equation}
   H(t)\Big|_{(N+3,-2,-N)} =
   \left. 2t + H + {H' \over H}\right|_{(N+2,0,-N-1)} \ .
\end{equation}
The result then follows upon making the appropriate identifications.
\qed

One can verify that this transformation $ T_1 T^{-1}_2 $ (and its inverse)
is the only nontrivial one which can map the Hamiltonian $ H $ into a rational
function of $ H $ and $ H' $. We have
\begin{equation}
   H\Big|_{(\alpha_0+1,\alpha_1-2,\alpha_2+1)} =
   T_1 T^{-1}_2 H\Big|_{(\alpha_0,\alpha_1,\alpha_2)} = 
   2t + H + {(1\m \alpha_1)H' \over H+\alpha_1 f_2} \ ,
\end{equation}
and this is rational if $ \alpha_1 = 1 $ (trivial case) or if $ \alpha_1 = 0 $
(this case). All other transformations are algebraic functions of $ H $ and
$ H' $.

\subsection{Calculation of $ F_N(s;a) $}
Turning our attention to $F_N$, we can make the following identifications.
\begin{proposition}
The average of the power of the characteristic polynomial is given by the 
$ N^{\text{th}}$ member of the $ \tau$-function sequence generated by the shift 
operator $ T^{-1}_3 $ from the initial parameters $ (1+a,-a,0) $,
\begin{equation}\label{bfq}
F_N(\lambda;a) = \bar{\tau}_3[N](\lambda;-a), 
\end{equation}
with the normalization of $ \bar{\tau}_3^{(N)}$ chosen so that 
\begin{equation}\label{nz}
F_N(\lambda;a) \sim \lambda^{Na} \quad {\rm as} \quad \lambda \to \infty \ ,
\end{equation}
(Eq.~(\ref{ent1})). 
The logarithmic derivative of the average is related to the Hamiltonian by
\begin{equation}
   V_N(t;a) = H\Big|_{(1+a+N,-a,-N)} \ ,
\label{V-Ham}
\end{equation}
(Eq.~(\ref{UH})).
\end{proposition}
\noindent {\it Proof}.
We see from (\ref{FN}) that
\begin{equation}\label{53a}
F_N(\lambda;a) = {1 \over C} 
\int_{-\infty}^\infty dx_1 \cdots \int_{-\infty}^\infty dx_N \,
\prod_{j=1}^N e^{-x_j^2} (\lambda - x_j)^a
\prod_{1 \le j < k \le N} (x_k - x_j)^2,
\end{equation}
where the normalization is such that (\ref{nz}) is satisfied.
Proceeding as in the derivation of (\ref{aw}) and (\ref{ent2})
we see that this can
be written in terms of determinants according to
\begin{equation}\label{53b}
F_N(\lambda;a)  =  {1 \over C}
\det \Big [ \int_{-\infty}^\infty (\lambda - x)^{a+j+k}
e^{-x^2} \, dx \Big ]_{j,k=0,\dots,N-1}.
\end{equation}
This is precisely the determinant occurring in (\ref{tu1'}) so we have
(\ref{bfq}).
Recalling (\ref{rmt3b}), we thus have
\begin{equation}\label{bf1}
F_N(\lambda;a) = F_N(\lambda_0;a)
	\exp \Big( \int_{\lambda_0}^\lambda V_N(t;a) \, dt \Big) \ , 
\end{equation}
where $V_N(t;a)$ is given in terms of $H$ with 
$(\alpha_0, \alpha_1, \alpha_2)$ equal to $(1+N+a,-a,-N)$ as in the formula 
(\ref{UH}) relating $U_N(t;a)$ to $H$. 
\qed

Thus $V_N(t;a)$, like $U_N(t;a)$ (recall (\ref{rsw})), satisfies the nonlinear
equation
\begin{equation}\label{rsw8}
(V_N'')^2 - 4 (tV_N' - V_N)^2 + 4V_N'(V_N' - 2a)(V_N'+2N) = 0. 
\end{equation}
The asymptotic behaviour (\ref{nz}) together with (\ref{bf1}) implies
(\ref{rsw8}) is to be solved subject to the boundary condition
\begin{equation}\label{rs3}
V_N(t;a) \sim {N a \over t}
\Big (1 + O(1/t) \Big ) \quad {\rm as} \quad t \to \infty.
\end{equation}
Apart from the quantity $R(t)$, which we know decays as a Gaussian, this
boundary condition is the same as (\ref{rsw1}). Thus $U_N(t;a)$ and
$V_N(t;a)$ satisfy the same differential equation, and up to a
term which decays as a Gaussian, the same boundary condition as
$t \to \infty$. However the $t \to - \infty$ behaviours are very different:
for $U_N(t;a)$ it is given by (\ref{rsw6}), while for $V_N(t;a)$ it is
(up to a possible phase) again given by (\ref{rs3}). 
In addition $V_N(t;a)$ satisfies the $ N-$difference equation (\ref{Udiff-T3})
and its $ a-$difference analogue but with the appropriate boundary conditions.

It was noted in the Introduction that for $(N\+ 1) \times (N\+ 1)$
dimensional GUE matrices the density, $\rho(\lambda)
\Big |_{N \mapsto N + 1}$ say, is proportional to
$e^{-\lambda^2} F_N(\lambda;2)$. Hence, analogous to (\ref{ag}), we see
from (\ref{bf1}) that
\begin{equation}\label{bf7}
\rho(\lambda) \Big |_{N \mapsto N + 1} = 
\rho(\lambda_0) \Big |_{N \mapsto N + 1}
\exp \Big( \int_{\lambda_0}^\lambda \left[ -2t + V_N(t;2) \right]\; dt \Big).
\end{equation}
On the other hand, we know $\rho(\lambda)
\Big |_{N \mapsto N + 1}$ is proportional to the $2 \times 2$ determinants
of (\ref{1.1}). In fact $F_N(\lambda;a)$, for general $a \in \Z_{>0}$ can
be written as an $a \times a$ determinant, a fact which can be understood
in the present setting by considering the $\tau$-function sequence
(\ref{tq}).

\begin{proposition}
The average of the powers of the characteristic polynomial $ F_N(\lambda;a) $ 
obey the duality relation
\begin{equation}\label{70}
{F_N(\lambda;a) \over F_N(\lambda_0;a)} =
{F_a(i\lambda;N) \over F_a(i\lambda_0;N)} \ ,
\end{equation}
for all $ a, N \in \Z $.
\end{proposition}
\noindent {\it Proof}.
First, note that by reversing the steps which led to
(\ref{aw}) and recalling (\ref{53a}) the determinant of (\ref{tq}) which
specifies $\bar{\tau}_1[n]$ can be written as a multiple integral to give
\begin{equation}\label{pe1}
\bar{\tau}_1[n](\lambda;\alpha_2) = C F_n(i\lambda; - \alpha_2) \ ,
\end{equation}
where $|C| = 1$. On the other hand, we see from (\ref{7.a})
that the Hamiltonian corresponding to $\bar{\tau}_1[n]$,
\begin{equation}
H \Big |_{(\alpha_0,\alpha_1,\alpha_2) =
(1 - \alpha_2 +n, -n, \alpha_2)},
\end{equation}
satisfies the differential equation (\ref{rsw}) with
\begin{equation}\label{pe2}
N = - \alpha_2, \qquad a = n.
\end{equation}
(Note that for the latter identification to be possible, we require
$a \in \Z_{\ge 0}$.) Furthermore, from (\ref{pe1}) and
(\ref{nz}), we have that
\begin{equation}\label{pe3}
{d \over dt} \log \bar{\tau}_1[n](t;\alpha_2)
\mathop{\sim}\limits_{t \to
\infty} - {n \alpha_2 \over t}\ ,
\end{equation}
which with the substitutions (\ref{pe2}) is identical to the boundary
condition of (\ref{rs3}). It follows from these facts that
\begin{equation}
F_a(i \lambda;N) = 
F_a(i \lambda_0;N) \exp \Big( \int_{\lambda_0}^\lambda V_N(t;a) \; dt \Big).
\end{equation}
Comparison with (\ref{bf1}) then yields (\ref{70}). 
\qed

From the definition (\ref{53a}) this identity implies the integral identity
\begin{align}
   & \int_{-\infty}^\infty dx_1 \cdots \int_{-\infty}^\infty dx_N
   \prod_{j=1}^N e^{-x_j^2} ( \lambda - x_j)^a
   \prod_{1 \le j < k \le N} (x_k - x_j)^2
   \nonumber \\
   & \qquad =
    C \int_{-\infty}^\infty dx_1 \cdots \int_{-\infty}^\infty dx_a
   \prod_{j=1}^a e^{-x_j^2} (\lambda - i x_j)^N
   \prod_{1 \le j < k \le a} (x_k - x_j)^2 \ ,
\label{60}
\end{align}
and from (\ref{53b}) gives the determinant identity
\begin{equation}\label{61}
\det \Big [ \int_{-\infty}^\infty (\lambda - x)^a x^{j+k}
e^{-x^2} \, dx \Big ]_{j,k=0,\dots,N-1} = C
\det \Big [ \int_{-\infty}^\infty (\lambda - i x)^N x^{j+k}
e^{-x^2} \, dx \Big ]_{j,k=0,\dots,a-1}.
\end{equation}
The integral identity (\ref{60}) has been derived earlier in the context
of a theory of generalized Hermite polynomials based on symmetric
Jack polynomials \cite{BF-97}, and in fact can be generalized so that
on the left-hand side the exponent 2 in the product of differences is
replaced by $2c$ and $ x^2_j \mapsto cx^2_j $ in the Gaussian, 
while on the right-hand side this same exponent is
replaced by $2/c$.  Regarding the determinant identity, noting that the
right-hand side is proportional to
\begin{equation}
\det \Big [ \int_{-\infty}^\infty (\lambda - ix)^{N+j+k}
e^{-x^2} \, dx \Big ]_{j,k=0,\dots,a-1} = C
\det \Big [ H_{N+j+k}(\lambda) \Big ]_{j,k=0,\dots,a-1}\ ,
\end{equation}
this gives a determinant formula for $ F_N(\lambda;a) $,
equivalent to that given by Br\'ezin and Hikami \cite{BH-99}.
Determinants with a Hankel structure constructed with orthogonal polynomial
elements are termed Tur\'anians and their positivity and other properties
such as relations with novel Wronskians have been investigated extensively, 
and reviewed in Karlin and Szeg{\"o} \cite{KS-82}. Explicit evaluations of 
Tur\'anians of the Hermite polynomials in terms of the Barnes $ G$-function 
where the initial degree of the polynomial is zero ($ N= 0 $) have been 
given by Radoux \cite{Ra-90}.

\subsection{$U_N(t;a)$ and $V_N(t;a)$ as Painlev\'e transcendents}
The formula (\ref{UH}) relates $U_N(t;a)$ to the Hamiltonian
(\ref{7.a}). Substituting the appropriate values of $\alpha_1$ and
$\alpha_2$ in the first equality of (\ref{7.a}) and recalling
(\ref{2.11'}) shows
\begin{equation}\label{fe}
U_N(t;a) = {1 \over 8q} (q')^2 - {1 \over 8} (q+2t)(q^2+2tq - 4a) -
{a^2 \over 2q} + Nq ,
\end{equation}
where $q$ satisfies the PIV equation (\ref{PIV}) with
\begin{equation}\label{fe1}
\alpha = 2N + 1 + a, \qquad \beta = - 2 a^2.
\end{equation}

In the case $a=0$ the functional expression (\ref{fe}) agrees with that
presented earlier \cite{TW-94,WFC-00a}, although the transcendent $q$ in the
earlier works is the PIV transcendent with $\alpha = 2N-1$, $\beta = 0$
rather than $\alpha = 2N + 1$, $\beta = 0$ as given by (\ref{fe1}). In fact
it follows from the work \cite{CS-93} that in general Eq. (\ref{f.5})
has more than one expression in terms of 
Painlev\'e transcendents. In the case $\alpha_1 = N$, $\alpha_2 = 0$ of this
equation the results of \cite{CS-93} give the functional expression implied
(\ref{fe}) with $q$ the PIV transcendent specified with the parameters 
$\beta = 0$ and either $\alpha = 2N + 1$ or $\alpha = 2N-1$ thus reconciling
(\ref{fe}) in the case $a=0$ with the results of \cite{TW-94,WFC-00a}.
We remark that the theory of \cite{CS-93} gives distinct functional forms
for the derivative,
\begin{equation}
U'_N(t;0) = - {1 \over 2} \epsilon\, q' - {1 \over 2} q^2 - tq,
\end{equation}
in the two case $\alpha = 2N + \epsilon$ ($\epsilon = \pm 1$).

Comparison of the formulas (\ref{V-Ham}) and (\ref{UH}) shows
$V_N(t;a)$ is given by the same Hamiltonian as $U_N(t;a)$. Thus
(\ref{fe}) remains true with the function $U_N(t;a)$ replaced by
$V_N(t;a)$ on the left-hand side.

\section{Edge Scaling in the GUE}
\setcounter{equation}{0}
\subsection{Calculation of $ E^{\rm soft}(s) $ and 
$ \tilde{E}^{\rm soft}(s;a) $}
To leading order the support of the eigenvalue density for $N \times N$
GUE matrices is the interval $(-\sqrt{2N}, \sqrt{2N})$. To study distributions
in the neighbourhood of the largest eigenvalue one shifts the origin to
the edge at $\sqrt{2N}$ 
and then scales the coordinate so
as to make the spacings of order unity in the $N \to \infty$ limit. This
is achieved by the mapping \cite{Fo-93}
\begin{equation}\label{4.1}
\lambda \mapsto \sqrt{2N} + {\lambda \over \sqrt{2} N^{1/6}}.
\end{equation}
Suppose we make this replacement (in the $s$-variable) in the probability
$E_N(0;(s,\infty))$ as specified by (\ref{ni2}). Then with
\begin{equation}
E^{\rm soft}(s) := \lim_{N \to \infty} E_N\Big (0;
\sqrt{2N} + {s \over \sqrt{2} N^{1/6}} \Big ) \ ,
\end{equation}
(because the eigenvalue density is not strictly zero outside the leading 
order of its support the edge is referred to as a soft edge) we see that
\begin{equation}\label{4.1a}
E^{\rm soft}(s) = \exp \Big( -\int_s^\infty r(t) \; dt \Big) \ ,
\end{equation}
where
\begin{equation}
r(t) = \lim_{N \to \infty} {1 \over \sqrt{2} N^{1/6}}
R \Big ( \sqrt{2N} + {t \over \sqrt{2}  N^{1/6}} \Big ).
\end{equation}
Furthermore, it follows from changing variables
$s \mapsto \sqrt{2N} + s/ \sqrt{2}  N^{1/6}$ in (\ref{c1a}),
replacing $R(\sqrt{2N} + s/ \sqrt{2}  N^{1/6})$ by
$ \sqrt{2}  N^{1/6} r(s)$ and taking the limit $N \to \infty$
that $r(s)$ satisfies the differential equation
\begin{equation}\label{4.2}
(r'')^2 + 4 r' \Big ( (r')^2 - sr' + r \Big ) = 0,
\end{equation}
a result first obtained by Tracy and Widom \cite{TW-94a}. Equation
(\ref{4.2}) is a particular case of the Jimbo-Miwa-Okamoto $\sigma$
form of the Painl\'eve II equation. We will find that the edge scaling is
essentially the second coalescence limit of the PIV system to the PII as
discussed in Subsect. 3.5.

\begin{proposition}
Define the scaling limit of the quantity $ \tilde{E}_N(s;a) $ by
\begin{equation}\label{4.3a}
\tilde{E}^{\rm soft}(s;a) :=
\lim_{N \to \infty} \Big ( C e^{-as^2/2} \tilde{E}_N(s;a) \Big )
\Big |_{s \to \sqrt{2N} + s/\sqrt{2} N^{1/6}} \ .
\end{equation}
Then
\begin{equation}\label{4.3b}
\tilde{E}^{\rm soft}(s;a) =
\tilde{E}^{\rm soft}(s_0;a) \exp \Big( \int_{s_0}^s u(t;a) \; dt \Big) \ ,
\end{equation}
where
\begin{align}
   u(t;a) 
   & = \lim_{N \to \infty} {1 \over \sqrt{2} N^{1/6}}
   \Big( -at + U_N(t;a) \Big) \Big|_{t \mapsto \sqrt{2N}+t/\sqrt{2}N^{1/6}}
   \ ,\label{4.3c}\\
   & = 
   -2^{1/3} H(-2^{1/3}t) \Big|_{(\alpha_0, \alpha_1) = (1-a, a)}
   \ ,\label{w10a}
\end{align}
with $H(t)$ is given by (\ref{Ham-PII}).
The function $ u(s;a) $ satisfies a second order second degree differential 
equation of the general Jimbo-Miwa-Okamoto $\sigma$ form of the 
Painlev\'e II equation
\begin{equation}\label{4.4}
(u'')^2 + 4 u' \Big ( (u')^2 - s u' + u \Big ) - a^2 = 0,
\end{equation}
subject to the boundary condition
\begin{equation}\label{es-bcU}
   u(s;a) \mathop{\sim}\limits_{s \to - \infty}
   \quarter s^2 + {4a^2\m 1 \over 8s} + {(4a^2\m 1)(4a^2\m 9) \over 64s^4}
   + \ldots \ .
\end{equation}
The function $ u(s;a) $ also satisfies the third order difference equation, 
related to the alternate discrete Painlev\'e a-dPI equation,
equation
\begin{equation}                                 
   {a \over u(s;a\+ 1)-u(s;a\m 1)} + {a\+ 1 \over u(s;a\+ 2)-u(s;a)}
  =   s - \left[ u(s;a\+ 1)-u(s;a) \right]^2 \ .
\label{es-udiff}
\end{equation}                                                                  
\end{proposition}
\noindent {\it Proof}.
Unlike the probability $E_N(0;(s,\infty))$, we do not expect the soft edge 
scaling limit of the quantities $\tilde{E}_N(s;a)$ as specified by (\ref{ent})
to be well defined. For example, in the case $a=2$, it is the combination 
$ e^{-s^2} \tilde{E}_N(s;2) $ which is proportional to
$ p_{\rm max}(s) $, and thus which should have a well defined scaling limit.
This suggests that for general $a$ we consider the scaling limit of
\begin{equation}\label{4.34'}
   C e^{-as^2/2} \tilde{E}_N(s;a).
\end{equation}
According to (\ref{ent4}) we have
\begin{equation}\label{4.3}
   e^{-as^2/2} \tilde{E}_N(s;a) =
   e^{-as^2_0/2} \tilde{E}_N(s_0;a)
   \exp \Big( \int_{s_0}^s\left[ -at + U_N(t;a) \right] \; dt \Big).
\end{equation}
Relation (\ref{4.3c}) follows from the definitions of 
$ \tilde{E}^{\rm soft}(s;a) $ and $ u(t;a) $ and (\ref{4.3}). The scaling 
in (\ref{4.3c}) is identical to the coalescence of the PIV system to the PII 
defined in Prop. 20, with the identifications 
$ \epsilon = (2N)^{-1/6} $, $ \alpha_{(IV)1} = -a $ and (\ref{UH}) for the 
relationship of the PIV Hamiltonian and $ U_N(t;a) $, and the scale changes 
$ t_{\rm II} = -2^{1/3}t $ and 
$ H_{\rm II}(t_{\rm II})\Big|_{(\alpha_0,\alpha_1)=(1-a,a)} = -2^{-1/3}u(t;a) $.
Proceeding as in the derivation of (\ref{4.2}) we find from (\ref{rsw}), 
or from (\ref{PII-Hode}) using the change of scale (\ref{w10a}), that $ u $ 
satisfies (\ref{4.4}).

To formulate the boundary condition for $u(t;a)$, we first recall
\cite{TW-94a} that the
$s \to -\infty$
boundary condition of $r(s)$ in (\ref{4.2}) is given by 
\begin{equation}\label{4.5}
r(s) \mathop{\sim}\limits_{s \to - \infty}
 \quarter s^2 - {1 \over 8 s} + {9 \over 64 s^4} + O\Big( {1 \over s^7} \Big),
\end{equation}
and this corresponds to the asymptotic behaviour \cite{TW-94a}
\begin{equation}\label{4.5a}
E^{\rm soft}(s) \mathop{\sim}\limits_{s \to - \infty}
\exp \Big( {s^3 \over 12} + {1 \over 8} \log (-s) \Big).
\end{equation}
Also, we know that $\tilde{E}_{\rm soft}(s;2)$ is proportional to the
derivative of $E^{\rm soft}(s)$, which implies
\begin{equation}\label{4.5b}
\tilde{E}_{\rm soft}(s;2) \mathop{\sim}\limits_{s \to - \infty}
\exp \Big( {s^3 \over 12} + C \log (-s) \Big).
\end{equation}
This suggests that for general $a$ we seek a solution of (\ref{4.4})
with the $s \to - \infty$ boundary condition
\begin{equation}\label{4.5c}
u(s;a) \mathop{\sim}\limits_{s \to - \infty}
{1 \over 4} s^2 + \sum_{j=1}^\infty {c_j \over s^j} .
\end{equation}
Substitution shows that in fact (\ref{4.4}) has a unique solution of
this form, and furthermore
\begin{equation}\label{4.5d}
c_1 = {4a^2-1 \over 8}, \quad c_2 = c_3 = 0, \quad
c_4 = {(4a^2-1)(4a^2-9) \over 64}, \quad \dots .
\end{equation}
The third order difference equation is just (\ref{HdiffE-PII}) with the scale 
change (\ref{w10a}).
\qed

We see from (\ref{es-bcU}) that the asymptotic expansion of $ u(s;a) $
terminates for $ a $ equal to half an odd integer. Recalling (\ref{w10a}) 
this is the case of $\alpha_1$ half an odd integer of the PII Hamiltonian 
(\ref{Ham-PII}). From the text about (\ref{PII-rational}) we know that this 
is precisely the parameter value for which the PII Hamiltonian can be 
expressed as a rational function of $ t $.

The fact that $e^{-s^2} \tilde{E}_N(s;2)$ is proportional to $p_{\rm max}(s)$
implies that the corresponding quantities in the scaled limit,
$\tilde{E}^{\rm soft}(s;2)$ and $p_{\rm max}^{\rm soft}(s)$
are proportional. Hence
\begin{equation}\label{pms}
p_{\rm max}^{\rm soft}(s) =
p_{\rm max}^{\rm soft}(s_0)
\exp \Big( \int_{s_0}^s u(t;2) \; dt \Big).
\end{equation}
The relation
\begin{equation}
p_{\rm max}^{\rm soft}(s) = {d \over ds} E^{\rm soft}(s)
\end{equation}
(Eq.~(\ref{1.1'})) with $E^{\rm soft}(s)$ specified by (\ref{4.1a})
then implies an identity between transcendents analogous to that
of Prop. \ref{pUR}.
\begin{proposition}
The quantity $u(t;2)$ of (\ref{pms}) and the quantity $r(t)$ of
(\ref{4.1a}) are related by
\begin{equation}\label{fy}
u(t;2) = {d \over dt} \log r(t) + r(t).
\end{equation}
\end{proposition}

\noindent {\it Proof}.
With
\begin{equation}
H(t) \Big |_{\alpha_1 = n} := H[n] ,
\end{equation}
it follows from (\ref{w10a}) and the substitution $r(t) = u(t;0)$ that
(\ref{fy}) is equivalent to
\begin{equation}
H[2] = {d \over dt} \log H[0] + H[0].
\end{equation}
But from the analogue of the first equality in (\ref{tau-PII}), together with 
(\ref{Bxfm-shift-PII:g}) and Table \ref{A1-Bxfm},
\begin{align*}
   H[2] = T_2^2 H[0] 
   & = \left. H[0] + q[0] + {\alpha_1 \over f_1[0]} 
            + T_2 \left( q[0] + {\alpha_1 \over f_1[0]} \right)
       \right|_{\alpha_1=0} \\
   & = H[0] + {1 \over 2 q^2[0] + t - p[0]}.
\end{align*}
Furthermore the first equality in (\ref{PII-derH}) together with
(\ref{Ham-PII}) in the case $\alpha_1 = 0$ give
\begin{equation}
{1 \over 2 q^2[0] + t - p[0]} = {d \over dt} \log H[0].
\end{equation}
\qed

\noindent {\it Note}.
According to (\ref{w10a}) we therefore have
\begin{equation}\label{w9}
   - 2^{-1/3} u( -2^{-1/3}t;a) =
   \half \left[ q'(t,a\m \half) \right]^2 -
   \half \left[ q^2(t,a\m \half) + {t \over 2} \right]^2
   - a q(t,a\m \half) ,
\end{equation}
where $q = q(t,\alpha)$ satisfies the PII equation (\ref{PII}). Also the 
first member of (\ref{PII-derH}), along with (\ref{coord-PII}) and 
(\ref{PII-Heqn}), gives
\begin{equation}\label{w10}
   -2^{-1/3} {d \over dt}u( -2^{-1/3}t;a)
   = -\half \left[ q'(t,a\m \half) + q^2(t,a\m \half) + {t \over 2} \right] .
\end{equation}
In the special case $a=0$ we see from (\ref{PII-ends}) that (\ref{w10})
simplifies to read
\begin{equation}\label{w11}
   u'(t;0) = - q^2(t,0)
\end{equation}
while (\ref{w9}) simplifies to read
\begin{equation}\label{w12}
   u(t;0) = \left[ q'(t,0) \right]^2 - t q^2(t,0) - q^4(t,0) .
\end{equation}
The results (\ref{w11}) and (\ref{w12}), deduced in a different way,
can be found in \cite{TW-94a}.  

\subsection{Calculation of $ {F}^{\rm soft}(\lambda;a) $}
Let us next consider the scaled quantity
\begin{equation}\label{4.4a}
{F}^{\rm soft}(\lambda;a) :=
\lim_{N \to \infty} \Big ( C e^{-a\lambda^2/2} {F}_N(\lambda;a) \Big )
\Big |_{\lambda \mapsto \sqrt{2N} + \lambda/\sqrt{2} N^{1/6}},
\end{equation}
where ${F}_N(\lambda;a)$ is specified by (\ref{53a}).
Because of the analogy with $\tilde{E}^{\rm soft}(s;a)$, which follows
from the identical structure of (\ref{ent4}), (\ref{rsw}) and
(\ref{bf1}), (\ref{rsw8}), analogous to (\ref{4.3b}) we have
\begin{equation}\label{4.4b}
   F^{\rm soft}(\lambda;a) :=
   F^{\rm soft}(\lambda_0;a)
   \exp \Big( \int_{\lambda_0}^\lambda v(t;a) \; dt \Big) \ ,
\end{equation}
where
\begin{equation}\label{4.3c'}
   v(t;a) = \lim_{N \to \infty} {1 \over \sqrt{2} N^{1/6}} \Big (
   -at + V_N(t;a) \Big ) \Big |_{t \mapsto \sqrt{2N} + t /\sqrt{2} N^{1/6}}
\end{equation}
satisfies the differential equation (\ref{4.4}) and the difference equation
(\ref{es-udiff}). The only difference between the logarithmic derivatives 
$v(t;a)$ and $u(t;a)$ is the boundary condition.

\begin{proposition}
The scaled averages of the powers of the characteristic polynomial 
$ F^{\rm soft}(\lambda;a) $ for $ a \in \Z_{\geq 0} $ have the determinantal 
form 
\begin{equation}\label{air}
F^{\rm soft}(\lambda;a) = (-1)^{a(a-1)/2}
\det \Big [ {d^{j+k} \over d\lambda^{j+k}} {\rm Ai} \, (\lambda)
\Big ]_{j,k=0,\dots,a-1} \ ,
\end{equation}
which was shown by Okamoto \cite{Ok-86} to be a $\tau$-function sequence for 
Painlev\'e II (recall Prop. 18). 
Furthermore this scaled average has a multiple integral representation of
the Kontsevich form \cite{Ko-92},
\begin{equation}\label{kn}
F^{\rm soft}(\lambda;a) = { (-1)^{a(a-1)/2} \over (-2 \pi i )^a}
\int_{-i\infty}^{i\infty} dv_1 \cdots \int_{-i\infty}^{i\infty}
dv_a \, \prod_{j=1}^a e^{v_j^3/3 - \lambda v_j}
\prod_{1 \le j < k \le a} (v_k - v_j)^2.
\end{equation}
The logarithmic derivative $ v(t;a) $ has the asymptotic expansion
\begin{equation}
  v(t;a)  \mathop{\sim}\limits_{t \to \infty}
  -a t^{1/2} - {a^2 \over 4t} + {a(4a^2\+ 1) \over 32t^{5/2}} \ .
\label{air-v}
\end{equation}
\end{proposition}
\noindent {\it Proof}.
For positive integer $a$ we can determine the $\lambda \to \infty$
behaviour of $F^{\rm soft}(\lambda;a)$, and thus the corresponding
behaviour of $v(t;a)$, by making use of the scaled form of the right-hand 
side of (\ref{53a}). To determine this scaled form, we first
require the explicit values of the constants in (\ref{53a}), (\ref{61})
and (\ref{4.4a}). Let us denote these constants by $C_1$, $C_2$,
$C_3$ respectively. Then from (\ref{53a}) and (\ref{nz}) we read off
that
\begin{align}
   C_1 \: := c(N)
   & = \int_{-\infty}^\infty dx_1 \cdots \int_{-\infty}^\infty dx_N \,
     \prod_{j=1}^N e^{-x_j^2} \prod_{1 \le j < k \le N} (x_k - x_j)^2
  \nonumber \\
   & = 2^{-N^2/2} (2 \pi )^{N/2} G(N\+ 2),
\label{cef1}
\end{align}
where $G(x)$ denotes the Barnes $G$-function, characterised for $x$
a positive integer by the functional property $G(x+1) = \Gamma(x) G(x)$
and the initial value $G(1) = 1$. The integral evaluation in (\ref{cef1})
can be derived by making use of the Vandermonde determinant identity
(\ref{van}) written in terms of Hermite polynomials. 

The proportionality constant $C$ in (\ref{61}) is the same as that in
(\ref{60}) and thus from (\ref{cef1}) given by
\begin{equation}\label{cef2}
C_2 = {c(N) \over c(a)}.
\end{equation}
Finally, we seek the value of the constant in (\ref{4.4a}). We know that
in the case $a=2$, $e^{-a\lambda^2/2} F_N(\lambda;a)$ is proportional to
the eigenvalue density for $(N\+ 1) \times (N\+ 1)$ dimensional GUE matrices.
Specifically
\begin{equation}
\rho(\lambda) \Big |_{N \mapsto N+1} =
(N\+ 1) {c(N) \over c(N\+ 1)} e^{-\lambda^2} F_N(\lambda;2).
\end{equation}
Since it is the combination $\rho(\lambda) \, d \lambda$ which has a scaled
limit, it follows that in the case $a=2$, $C_3 = (N\+ 1) c(N)/
(c(N\+ 1) \sqrt{2} N^{1/6})$. This suggests that for general $a \in \Z_{\ge 0}$
we should choose
\begin{equation}\label{cef3}
C_3 = c_a {\Gamma (N\+ 1\+ a/2) \over \Gamma (N\+ 1)}
{c(N) \over c(N\+ a/2)} \Big ( {1 \over \sqrt{2} N^{1/6}} \Big )^p \ ,
\end{equation}
where $p$ is a power to be determined and $c_a$ depends only on $a$
($p=1, c_a=1$ for $a=2$). 

Substituting (\ref{cef1}), (\ref{cef2}), (\ref{cef3}) in (\ref{53a}),
(\ref{61}), (\ref{4.4a}) respectively we see that for $a \in \Z_{>0}$,
\begin{align}
   F^{\rm soft}(\lambda;a)
   & = c_a \lim_{N \to \infty}
     {\Gamma (N\+ 1\+ a/2) \over \Gamma (N\+ 1)} 
     {c(N) \over c(N\+ a/2) c(a)} \Big( {1 \over \sqrt{2}N^{1/6}} \Big)^p
     e^{-a \lambda^2/2}
   \nonumber \\
   & \times \det \Big[
     \int_{-\infty}^\infty (\lambda - ix)^N x^{j+k} e^{-x^2} \, dx
     \Big|_{\lambda \mapsto \sqrt{2N} + \lambda/\sqrt{2} N^{1/6} }
     \Big]_{j,k=0,\dots,a-1}.
\label{ne1}
\end{align}
But analogous to the equality in (\ref{aw}) we have
\begin{equation}\label{aw5}
\det \Big [ \int_{-\infty}^\infty (\lambda - ix)^N
x^{j+k} e^{-x^2} \, dx  \Big ]_{j,k=0,\dots,a-1} =
(-1)^{a(a-1)/2} 
\det
\Big [  \int_{-\infty}^\infty (\lambda - ix)^{N+j+k} e^{-x^2} \, dx
\Big ]_{j,k=0,\dots,a-1}.
\end{equation}
This can be further rewritten by noting that analogous to
(\ref{an1}),
\begin{align}
   \int_{-\infty}^\infty (\lambda - ix)^{N+j+k} e^{-x^2} \, dx
   & = (-2)^{-(j+k)} e^{\lambda^2} 
     {d^{j+k} \over d\lambda^{j+k}}
     \Big( e^{-\lambda^2} \int_{-\infty}^\infty
     (\lambda - i x)^N e^{-x^2} \, dx \Big)
   \nonumber \\
   & = (-2)^{-(j+k)} e^{\lambda^2} 2^{-N} \sqrt{\pi}
     {d^{j+k} \over d\lambda^{j+k}}
    \Big( e^{-\lambda^2} H_N(\lambda) \Big). 
\label{ne}
\end{align}

Making use of the asymptotic expansion for the Barnes $G$-function
\cite{Ba-1900}
\begin{equation}\label{barnes}
\log G(x+1) \mathop{\sim}\limits_{x \to \infty}
 {x^2 \over 2} \log x - {3 \over 4} x^2 + {x \over 2}
\log 2 \pi -{1 \over 12} \log x + {\rm O}(1),
\end{equation}
and the Plancherel-Rotach asymptotic expansion of the Hermite polynomials 
\cite{ops-Sz}
\begin{equation}\label{7.her}
\exp(-x^2/2) H_N(x) = 
	\pi^{-3/4}2^{N/2+1/4}(N!)^{1/2}N^{-1/12}\{\pi\mbox{Ai}(-t/3^{1/3}) 
	+ {\rm O}(N^{-2/3})\} \ ,
\end{equation}
where $x=(2N)^{1/2} - 2^{-1/2}3^{-1/3}N^{-1/6}t$ and with Ai$(x)$ denoting the
Airy function,
we see from Eqs. (\ref{ne1}), (\ref{aw5}) and (\ref{ne})
that with $p = a/2$ in (\ref{cef3}) and 
appropriate $c_a$, the determinantal representation (\ref{air}) holds.
Furthermore, in the case $a=2$ we read off the functional form
\begin{equation}\label{air1}
\Big ( {\rm Ai}'(x) \Big )^2 - {\rm Ai}(x) {\rm Ai}''(x) \ ,
\end{equation}
which is the known expression \cite{Fo-93} for the scaled soft edge
density in the GUE. Another point of interest, which follows from the
integral formula
\begin{equation}\label{air7}
{\rm Ai}(x) = {1 \over 2 \pi i} \int_{-i\infty}^{i\infty}
	\exp \Big( {1 \over 3} v^3 - xv \Big) \, dv,
\end{equation}
is that (\ref{air}) can be written
\begin{equation}\label{air8}
F^{\rm soft}(\lambda;a) = { (-1)^{a(a-1)/2} \over (-2 \pi i )^a}
\det \Big [ \int_{-i\infty}^{i\infty}
	\exp  \Big( {1 \over 3} v^3 - \lambda v \Big) v^{j+k} \, dv
     \Big ]_{j,k=0,\dots,a-1}  \ .
\end{equation}
Thus, reversing the reasoning leading from (\ref{3.0}) to (\ref{aw})
we have the multiple integral representation (\ref{kn}) for 
$ F^{\rm soft}(\lambda;a) $, which is an example of the class of integrals
studied by Kontsevich \cite{Ko-92}.

Consider now the asymptotic form of (\ref{air}). In the case $a=1$
this is just the Airy function, which has the known
$x \to \infty$ asymptotic form (see e.g.~\cite{asy-Ol} p.~116)
\begin{equation}\label{air2}
{\rm Ai} \,(x) \mathop{\sim}\limits_{x \to \infty}
{e^{-\xi} \over 2 \pi^{1/2} x^{1/4}}
\sum_{k=0}^\infty (-1)^k {u_k \over \xi^k} \ ,
\end{equation}
where $\xi := {2 \over 3} x^{3/2}$, $u_0 = 1$ and
\begin{equation}
u_k = {(2k+1) (2k+3) \cdots (6k-1) \over (216)^k k!}, \quad k \ge 1.
\end{equation}
It follows from this and (\ref{air}) that for general $a \in Z_{>0}$,
\begin{equation}\label{air3}
\log F^{\rm soft}(\lambda;a) \mathop{\sim}\limits_{\lambda \to \infty}
- {2 a \over 3} \lambda^{3/2} + C \log \lambda + c_0
+ \sum_{j=1}^\infty {\tilde{c}_j \over \lambda^{3j/2}},
\end{equation}
which in combination with (\ref{4.4b}) implies that we must seek
a solution of (\ref{4.4}) (with $u$ replaced by $v$) subject to the
boundary condition
\begin{equation}\label{air4}
v(t;a) \mathop{\sim}\limits_{t \to \infty}
- a t^{1/2} + {C \over t} +
\sum_{j=1}^\infty {c_j \over t^{3j/2+1}}.
\end{equation}
Substitution of (\ref{air4}) in (\ref{4.4}) shows there is a unique
solution of this form, with
\begin{equation}\label{air5}
C= - {a^2 \over 4}, \quad
c_1 = {a (1\+ 4a^2) \over 32}, \quad \dots
\end{equation}
given by (\ref{air-v}).
\qed

\section{Conclusions - A Programme}
\setcounter{equation}{0}
We have applied the Okamoto $\tau$-function theory of PIV and PII to the 
computation of $ \tilde{E}_N(s;a) $ and $ F_N(s;a) $ for the GUE and its 
scaled soft edge limit. As noted in the Introduction, the Okamoto 
$ \tau$-function theory applies equally as well to the computation of 
$ \tilde{E}_N(s;a) $ and $ F_N(s;a) $ for all matrix ensembles with a unitary 
symmetry and classical weight functions (\ref{weights}). Thus we expect to 
be able to compute $ \tilde{E}_N(s;a) $ and $ F_N(s;a) $ in the cases of the 
Laguerre, Jacobi and Cauchy ensembles (special cases of $ F_N(s;a) $ have been
evaluated in terms of Painlev\'e transcendents for the Laguerre ensemble
\cite{TW-99b}, and for the Jacobi ensemble \cite{AV-99}). In future studies 
we will undertake this task by following the programme used here for the GUE, 
the main steps of which can be itemised as follows:
\begin{itemize}
  \item
   From the definitions of the gap probability $E_N(0;I)$, $I$ a single
   interval including the boundary of the eigenvalue support, and
   $\tilde{E}_N(s;a), F_N(s;a) $ as $ N$-dimensional multidimensional
   integrals they can be converted into $ N\times N $ determinants
   analogous to (\ref{aw}), (\ref{ent2}) and (\ref{53b}) respectively.
  \item
   Using an identity analogous to (\ref{an1}), the determinants can be 
   put into the double Wronskian form (\ref{11.b}), with $d/dt$ replaced
   by
   \begin{equation}\label{dt}
      t{d \over dt}, \quad t(1 - t){d \over dt}
   \end{equation}
   in the Laguerre and Jacobi ensembles respectively.
  \item
   The Okamoto $\tau$-function theory of PV and PVI \cite{Ok-85,Ok-87a} 
   gives these same determinants as $\tau$-function sequences, in which 
   the initial members are $\tau[0] = 1$, and $\tau[1]$ the solution of 
   the particular classical equation associated with relevant Painlev\'e 
   transcendent when the parameter sequences begin on a wall of the Weyl
   chamber in the affine space of parameters. The classical solutions,
   and their polynomial specialisations, are noted for each of the
   Painlev\'e transcendents in Table \ref{P-classical},

  \begin{table}[H]
  \begin{center}
  \begin{tabular}{|c|c|c|}
  \hline

  PJ & Classical Solution & Classical \\
     &                    & Orthogonal Polynomial \\
  \hline
  PI & - & - \\
  \hline

  PII & Airy & - \\
  \hline

  PIII & Bessel & - \\
  \hline

  PIV & Hermite-Weber & Hermite \\
  \hline

  PV & Confluent Hypergeometric & Laguerre \\
  \hline

  PVI & Gau{\ss} Hypergeometric & Jacobi \\
  \hline

  \end{tabular}
  \caption{Classical solutions of the Painlev\'e transcendents.}
   \label{P-classical}
  \end{center}
  \end{table}

  \item
   The logarithmic derivatives (with $d/dt$ replaced by (\ref{dt}) as
   appropriate) $ R_N(s), U_N(s), V_N(s) $ coincide with the Hamiltonians
   in the Painlev\'e theory and as such satisfy certain second order 
   second degree ODEs of the Painlev\'e type.
  \item
   The $\tau$-function sequence $ \{ \tau_0[N](t;a)\}_{N \geq 0} $, say 
   corresponding to $F_N(s;a)$, is simply related to another $\tau$-function 
   sequence $ \{ \tau_1[a](t;N)\}_{a \geq 0 } $. Both $\tau$-functions 
   relate to the same Hamiltonian but result from the action of different
   shift operators. Because the shifts are commutative one has
  \begin{equation}
     {\tau_0[N](t;a) \over \tau_0[N](t_0;a)} =
     {\tau_1[a](t;N) \over \tau_1[a](t_0;N)} \ .
  \end{equation}
   Identities of this type for the Laguerre and Jacobi ensembles, written 
   as multiple integrals, are already known from \cite{BF-97}.
  \item
   For all the independent shift operators and sequences of
   $ q[n], p[n], H[n], \tau[n] $ there exist difference equations generated
   by the B\"acklund transformations of these shifts. It has been conjectured
   that all the difference equations arising in this way are discrete
   Painlev\'e equations satisfying
   integrability criteria such as singularity confinement analogous to the
   Painlev\'e criteria.
\item
   In the appropriate edge scaling limit, the analogues of $r(s), u(s), v(s)$
   are Hamiltonian functions for PII or PIII, and satisfy the corresponding
   second order second degree equation.
\end{itemize}

\begin{acknowledgement}
This research has been supported by the Australian Research Council.
PJF thanks M.~Noumi for explaining aspects of his work with Y.~Yamada,
and thanks K.~Aomoto for obtaining funds for his visit to Japan
in June 2000 which made that possible.
\end{acknowledgement}

\addcontentsline{toc}{section}{\numberline{}References}
\bibliographystyle{cmp}
\bibliography{moment,random_matrices,nonlinear}

\end{document}